\documentclass[aip,preprint]{revtex4-1}
\usepackage{amsmath}
\usepackage{setspace}
\usepackage[hidelinks,colorlinks=true,linkcolor=blue,citecolor=blue]{hyperref}
\usepackage{graphicx}
\usepackage{csquotes}

\draft % marks overfull lines with a black rule on the right

\begin{document}
\title{TUNABILITY OF DISSIPATIVE PARTICLE DYNAMICS SIMULATIONS FOR EXCLUDED VOLUME AND HYDRODYNAMIC INTERACTIONS IN POLYMER SOLUTIONS AND RHEOLOGICAL PREDICTIONS}

%List abbreviations here, if any. Please note that it is preferred that abbreviations be defined at the first instance they appear in the text, rather than creating an abbreviations list.
%\abbrevs{ABC, a black cat; DEF, doesn't ever fret; GHI, goes home immediately.}

%Include full author names and degrees, when required by the journal.
% Use the \authfn to add symbols for additional footnotes and present addresses, if any. Usually start with 1 for notes about author contributions, then continue with 2, etc if any author has a different present address.
\author{Sanjay Jana}
\author{Venkata Siva Krishna }%
 \affiliation{Department of Chemical Engineering, Indian Institute of Technology Kanpur, Kanpur-208016, India}%Lines break automatically or can be forced with \\
\author{Praphul Kumar }%
% \email{bharatk@nvidia.com}
\affiliation{Department of Chemical Engineering, Indian Institute of Technology Bombay, Maharashtra-400075, India%\\This line break forced with \textbackslash\textbackslash
}
\author{Indranil Saha Dalal}
 \email{indrasd@iitk.ac.in}
 \affiliation{Department of Chemical Engineering, Indian Institute of Technology Kanpur, Kanpur-208016, India}

%\author{}
%\thanks{}
%\affiliation{}


\date{\today}
%\affil[1]{Department of Chemical Engineering, Indian Institute of Technology, Kanpur, UP, 208016, India}
%\affil[2]{Department of Chemical Engineering, Indian Institute of Technology, Kanpur, UP, 208016, India}

%\corraddress{Indranil Saha Dalal,Department of Chemical Engineering, Indian Institute of Technology, Kanpur, UP, 208016, India}
%\corremail{indrasd@iitk.ac.in}

%\presentadd[\authfn{2}]{Department, Institution, City, State or Province, Postal Code, Country}

%\fundinginfo{Funder One, Funder One Department, Grant/Award Number: 123456, 123457 and 123458; Funder Two, Funder Two Department, Grant/Award Number: 123459}

%Include the name of the author, that should appear in the running header
%\runningauthor{Jana et al.}

%\begin{frontmatter}
%\maketitle
%\end{frontmatter}
\begin{abstract}
Even though the Dissipative Particle Dynamics (DPD) has shown its worth in a variety of research areas, it has been rarely used for polymer dynamics, particularly in dilute and semi-dilute conditions and under imposed flow
fields. For such applications, the most popular technique has been
Brownian dynamics (BD), even though the formulation of the same may be
complicated for flow in complex geometries, which is straightforward for
DPD. This is partly due to the flexibility of BD simulations to mimic any
dynamic regime for polymer solutions by independently tuning hydrodynamic
interactions (HI) and excluded volume (EV). In this study, we reveal that
DPD also offers a similar flexibility and the regimes with respect to
dominant EV and HI can be selected as conveniently as BD. This flexibility
is achieved by tuning the repulsive interaction parameter of polymer beads
and the spring length (which determines the chain resolution). Our results
show that the former sets the chain size (and thus, EV) while the latter
can be used to set the HI, nearly independently of each other. Thus, any
rheological regime of certain level of EV and HI can be attained by
appropriately tuning only these two parameters, providing a flexibility of
similar levels as BD simulations. We further indicate the suitability of
DPD by comparing rheological predictions with equivalent models in BD. For
this, we imposed startup uniaxial extensional flows and steady shear flows
on the system. Our results indicate the consistency of DPD with BD
simulations, which is known to agree well with experiments.
\end{abstract}

\maketitle
\section{Introduction}

The rheology of polymer solutions and the physics of macromolecular chains in flow fields have enormous relevance for industry as well as biological processes. Thus, they have been the subject of investigation for several decades, which includes theory, experiments, and simulations. In particular, the rheologically important quantities like stress and viscosity are linked to the conformational changes of the polymer chains in flow. The first significant step towards modeling of polymer chain dynamics was proposed by Rouse \cite{rouse1953theory}, who showed that the chain dynamics and relaxation processes can be decomposed into a hierarchy of modes, at equilibrium. For this, Rouse considered a simple model of beads connected by Hookean springs for the polymer chains immersed in a continuum solvent. Later, Zimm \cite{zimm1956dynamics} incorporated hydrodynamic interactions (HI) in the same model in a pre-averaged manner. However, the resulting model predicted scaling laws that agreed well with rheological measurements. Starting with this foundation, researchers have mostly used Brownian dynamics (BD) simulations for polymer solutions, especially for flow problems \cite{hsieh2003modeling, hsieh2004modeling, petera1999brownian, hur2000brownian}. Technological advances in CPUs enabled the relaxation of some assumptions in the Zimm model. Firstly, the Hookean springs are replaced by finitely extensible counterparts, most notably the ones using
Cohen-Padé approximation that closely mimics the inverse Langevin function. Secondly, the fluctuating HI (and not pre-averaged) can be computed, mostly through the Rotne-Prager-Yamakawa (RPY) tensor, to account for HI. 
 Additionally, a large ensemble of chains, even to the extent of a few hundreds, can be considered, especially for simulations under flow fields. All of these details are discussed in a review article \cite{larson2005rheology} and in a series of articles using BD simulations \cite{dalal2012multiple, kumar2022fraenkel, kumar2023effects}.

Despite nearly three decades of polymer dynamics simulations by primarily BD simulations, several issues have been noted that remain unexplained by BD simulations alone. The accuracy of BD simulations, particularly with HI (and the RPY tensor), has been under scanner in several studies that compared with experiments \cite{larson2005rheology, hsieh2003modeling, hsieh2004modeling}. Thus, even though many trends show good agreement for polymer chain simulations in flow fields, BD may not be the most accurate possible choice for such investigations, particularly with HI. A highly accurate alternative, with all details included, can be MD simulations that retains all atoms on the chain and solvent bath. However, MD for practically sized chains (about thousands to tens of thousands of Kuhn steps) is computationally prohibitive even with modern processors. This is owing to the large number of atoms (several millions at least) that need to be simulated for long time scales (several relaxation times, say for several milliseconds). BD provides a much cheaper alternative for such applications, with a maximum requirement of about tens of thousands of beads, even for a highly resolved chain model. This is since BD works in the mesoscale (and not atomistic), where every Kuhn step can be represented by one or more beads, or several Kuhn steps can be mimicked by one spring (or bead). Note, due to the massive computational cost, MD has been used for polymer dynamics simulations for very short chains (few Kuhn steps long i.e. few tens of monomers) in a few earlier studies \cite{saha2013explaining}.

Thus, for such studies, mesoscale simulations are preferable over atomistic ones for computational reasons. Nearly all earlier mesoscale simulations of polymer chains used BD, as discussed earlier. Another promising mesoscale simulation technique that can be used for such simulations is Dissipative Particle Dynamics (DPD). Before discussing the need of this study using DPD, let us summarize the merits and demerits of BD. The most obvious advantage is the computational cost, since the solvent phase is represented by a continuum. However, with HI, the cost scales as $~N^{3}$ due to a Cholesky decomposition at every timestep \cite{krishna2024analysis}. Thus, beyond 100 beads, BD simulations also become costly with HI. The second, and most important advantage, is the flexibility in modeling various polymer dynamics regimes. Apart from connectivity, the mechanisms of excluded volume (EV) and HI are either dominant or negligible for any given polymer solution. This gives rise to various dynamic regimes, depending on which mechanisms are dominant. In BD simulations, the mechanisms of EV and HI can be tuned independently of one another. Each has parameters of their own, that can be set independently, or even switched on and off. This provides a high degree of flexibility to traverse between various rheological regimes of polymer dynamics. Thus, the rheological studies of any given polymer solution can be performed using BD by setting the required levels of EV and HI, without any other tweaks. Among the demerits, BD simulations have been known to be not accurate for some rheological studies, as discussed earlier \cite{larson2005rheology, hsieh2003modeling, hsieh2004modeling}. Researchers doubt the HI formulation in BD, especially the RPY tensor, to explain differences with experiments. The most critical disadvantage of BD is regarding its usage in flow applications, especially in complicated geometries. Note, the RPY tensor is formulated for unbounded systems. Typical flow applications would involve chains in flow fields in bounded geometries. For simple geometries, there HI tensors may be formulated. However, these may be impossible to be obtained in complicated geometries.

Thus, from the applications perspective, a mesoscale technique is desired that retains the advantages of BD, while circumventing some of its issues. DPD is the other mesoscale alternative that is computationally feasible for similar problems. It has also been used for other problems involving complex fluids, such as to investigate the 3D flow dynamics of moving droplets \cite{li2013three} and simulate colloidal particle transport in confined and bulk geometries \cite{li2008hydrodynamic}. Other researchers have leveraged the potential of DPD to advance our understanding of multiphase fluid dynamics in microchannels \cite{liu2007dissipative}. In another study, researchers have investigated how wall elasticity influences hydrodynamic interactions, changing the drag force of the sphere and the flow field \cite{reddy2009dissipative}. In DPD, the HI is present implicitly, by exchange of momentum of polymer and solvent beads. Thus, no tensor is required to be formulated, and the method becomes generally applicable for any complex geometry with boundaries. This makes it highly desirable for flow problems. Surprisingly, there have been very few studies of polymer dynamics with DPD in solutions \cite{huang2006flow, wijmans2002simulating, jiang2007hydrodynamic}. Most of these investigated basic aspects of chain dynamics using beads and springs. Only one recent study used highly resolved polymer chains (to a single Kuhn step) in DPD and investigated aspects of both equilibrium and shear flow \cite{kumar2023effectiveness}. Using the default interaction potentials, this study showed that DPD does capture several experimentally (and through simulations as well) known trends for polymer dynamics. However, apart from the initial promise shown in this earlier study \cite{kumar2023effectiveness}, it is not clear if it can replace BD, or provide the same advantages as BD simulations. This is particularly about the flexibility aspect of BD simulations, where the dynamic regimes can be traversed by tuning the values for the mechanisms of EV and HI. Currently, despite showing success and also the potential ease of computing in complex geometrics, it is not clear if, and how, such tuning of EV and HI can be obtained in DPD. If such flexibility is possible, DPD may emerge as a better method than BD, from an overall perspective, for such problems.

Thus, in this study, we show that this flexibility or tunability of EV and HI is possible within DPD. That is, there exists parameters, through which we can set EV and HI levels, nearly independently of one another. Thus, any polymer dynamics regime, where EV or HI are dominant or negligible, can be modeled by suitably adjusting parameters in DPD simulations. As we show later, the soft repulsive potential between polymer beads control EV, while the spring length parameter controls HI, nearly independently of one another. For this study, a highly resolved chain is considered, to a single Kuhn step, as is the norm in recent BD \cite{kumar2024effects} and DPD \cite{kumar2023effectiveness} simulations. We also perform studies at both equilibrium and under rheologically important flow fields. These are compared with BD simulations, which is the most popular method for polymer dynamics and has been compared with experiments. Our results show consistency with known results for polymer dynamics as well as BD simulations in rheologically important flow fields.

\section{Simulation methodology}
DPD is an extremely popular technique in the mesoscale i.e. between atomistic and macroscopic length scales. As discussed, this simulation method has been used in a variety of situations for different systems. In DPD, several atoms are grouped together to form a \enquote{bead}. These beads interact via soft repulsive potentials. Apart from this, two more forces act on each bead - random and dissipative forces in a pairwise fashion. Thus, for each pair, one needs to compute three forces in DPD. Although DPD simulations have been discussed in detail in earlier studies \cite{kumar2023effectiveness}, we present a brief description in the following section.

\subsection{Mathematical formulation}

Let us consider a system of $N$ DPD beads, each having a 
mass $m$ for simplicity, with  position vectors $\vec{r}_i$ and velocity $\vec{v}_i$. The equation of motion of each bead follows from the Newton's second law of motion:
\begin{equation}\label{eq:1}  
m\frac{d\vec{v}_i}{dt} = \vec{F}_{ij}
\end{equation}
where $\vec{v}_i=d\vec{r}_i/{dt}$ and $\vec{F}_{ij}$ is the total  
force acting on the $i^{th}$ particle due to interactions with all other particles. The total force 
$\vec{F}_{ij}$ is calculated as : 
\begin{equation}\label{eq:2}
\vec{F}_{ij}=\vec{F}_{ij}^C +\vec{F}_{ij}^D +\vec{F}_{ij}^R
\end{equation}
where $\vec{F}_{ij}^C$, $\vec{F}_{ij}^D$ and $\vec{F}_{ij}^R$ denote the 
soft conservative, dissipative and random forces, respectively. These forces 
are pairwise additive and are given as \cite{groot1997dissipative} :
\begin{equation}\label{eq:3}
\vec{F}_{ij}^C = w^C (r_{ij}) \hat{r}_{ij}
\end{equation}
\begin{equation}\label{eq:4}
\vec{F}_{ij}^D = -\gamma w^D (r_{ij})(\hat{r}_{ij}\cdot 
\vec{v}_{ij})\hat{r}_{ij}
\end{equation}
\begin{equation}\label{eq:5}
\vec{F}_{ij}^R = \sigma w^R (r_{ij}) \theta_{ij} \hat{r}_{ij}
\end{equation}
where $ \vec{r}_{ij} = \vec{r}_i-\vec{r}_j$,
$\hat{r}_{ij}=\vec{r}_{ij}/|\vec{r}_{ij}|$,
and $ \vec{v}_{ij} = \vec{v}_i-\vec{v}_j$ are the relative position vector,  
corresponding unit vector and the relative velocity vector (of bead $i$  with respect to bead 
$j$), respectively. $w^C$, $w^D$ and $w^R$ denote the weight 
functions of the conservative, dissipative and random forces, respectively. 
The parameters $\gamma$ and $\sigma$  determine the magnitudes of the 
dissipative  and  random forces, respectively. The term $\theta_{ij}$ denote a random variable following a gaussian distribution with the symmetry property $ \theta_{ij}= \theta_{ji}$, which 
ensures the conservation of momentum. It has the following properties as \cite{groot1997dissipative} :

\begin{equation}\label{eq:6}
\left\langle  \theta_{ij}\right\rangle =0
\end{equation}
\begin{equation}\label{eq:7}
\left\langle  \theta_{ij}(t) \theta_{kl}(t')\right\rangle 
=(\delta_{ik}\delta_{jl}+\delta_{il}\delta_{jk})\delta(t-t')
\end{equation}

All the forces are assumed to act within a sphere of a cut-off radius $r_c$, which determines the 
length scale for all interactions. The weight function of the conservative force is given as :
\begin{equation}\label{eq:8}
w^C(r_{ij})  =
\begin{cases}
a_{ij}(1-r_{ij}/r_c)  &\text{if $ r_{ij}\leq r_c$}	\\
0 &\text{if $r_{ij}\geq r_c$}
\end{cases}
\end{equation}
where $a_{ij}$ is the repulsion parameter between beads $i$ and $j$. 
This $a_{ij}$ can be used to mimic various types of interactions. However, in most studies, the value of $a_{ij}$ is taken to be same for all similar type of beads. However, beads of different phases may have different $a_{ij}$ values. For consistency with the fluctuation-dissipation theorem, two 
conditions are set on the weight functions and amplitudes of the 
dissipative and random 	forces \cite{espanol1995statistical, groot1997dissipative}
\begin{equation}\label{eq:9}
w^D(r_{ij})  = [w^R(r_{ij})]^2
\end{equation} 
\begin{equation}\label{eq:10}
\sigma^2 = 2\gamma k_BT
\end{equation}
where $k_B$ is the Boltzmann constant and $T$ is the system temperature. 
In the standard DPD method, the weight function takes the following form \cite{groot1997dissipative}
\begin{equation}\label{eq:11}
w^R(r_{ij})  =
\begin{cases}
(1-r_{ij}/r_c)  &\text{if $ r_{ij}\leq r_c$}	\\
0 &\text{if $r_{ij}\geq r_c$}
\end{cases}
\end{equation}

In this study, all DPD simulations are performed using LAMMPS  \cite{plimpton1995fast}, unless otherwise mentioned.

\subsection{System and Selection of parameters}
In this study, we performed simulations within a cubic periodic box. Similar to the previous study \cite{kumar2023effectiveness}, a single polymer chain is immersed in a solvent bath. Thus, most of the setup and parameters closely follow the previous study \cite{kumar2023effectiveness}. The dimensions of the simulation box are sufficiently large to ensure that the equilibrium size of the polymer chain remained unaffected. The bead mass $(m)$, cut-off distance $(r_c)$, and thermal energy $(k_BT)$, are all set to unity. The friction coefficient $\gamma$ is set to a value of 4.5.
The number density $n=3$ is fixed for all simulations presented here. Three different $a_{ij}$ values are used for the three different type of interactions. These are denoted as $a_{ss}, a_{sp}$, $a_{pp}$, where the subscripts $p$ and $s$ referred to the polymer and solvent beads, respectively. Later, we indicate the values used for these parameters. Note, the default value used for $a_{ij}$ in most DPD simulations in literature is 25, which has also been used here and in the earlier study, as indicated later. Following the recent articles, we have used a bead-rod description for the polymer chain \cite{kumar2023effectiveness}. For this, as described in previous study \cite{kumar2023effectiveness}, two adjacent beads on the chain are connected by a harmonic bond. This bond is characterized by a potential energy function $E$, as follows:

\begin{equation}\label{eq:17}
E=K(r-b)^2    
\end{equation}

In our setup, the parameter $b$ signifies the equilibrium bond distance, while $K$ represents the spring constant, including the standard factor of $1/2$. Physically, $b$ denotes the length of a Kuhn step, relative to the interaction length scale ($r_{c}$). Thus, for a sufficiently stiff spring, the polymer chain is resolved to a single kuhn step.

As discussed earlier, we aim to investigate the tunability of DPD simulations in terms of accessing the different regimes of polymer dynamics, which are of enormous rheological interest. In the previous study \cite{kumar2023effectiveness}, we established the accuracy of DPD simulations in being able to capture certain aspects of polymer dynamics (with a bead-rod chain model, resolved to a single Kuhn step). However, we selected a set of parameters that simulated the regime where both HI and EV are dominant. It was observed that the resulting dynamics captured some trends observed from experiments as well.
However, the question of tunability, flexibility and accuracy of DPD to access different rheologically important regimes has never been investigated, to the best of our knowledge. Additionally, the appropriateness of DPD predictions in rheological flow fields in these different regimes has never been investigated. In this study, keeping the default parameter values to be same as our earlier study \cite{kumar2023effectiveness}, we attempt to further establish the appropriateness of DPD for polymer rheology. Additionally, we show how various regimes of polymer dynamics, with EV and HI being independently tunable, can be achieved through these parameters in DPD. Here, we briefly summarize the default values: $a_{ij}=25$ is used for all interactions, unless specified otherwise (when $a_{pp}$ is set independently), $K=5000$ except for investigations to observe the effects of spring stiffness. $b=0.85$ denotes the default length of one spring, (note: length scale is set by $r_{c}=1$), unless we study the effect of Kuhn length (and highlight its usefulness in terms of dynamical regimes). Note, the stiff, nearly inextensible springs mimic one Kuhn step.

\subsection{Chain Size and Relaxation time measures}
Here, we define the various quantities used as measures of chain size and dynamics. Note, these are usual definitions also used in several earlier articles \cite{doi1988theory}.

End-to-end distance: This is the magnitude of the vector connecting the terminal beads of the chain.
The RMS value is typically used, given as :
\begin{equation}
R_{end-to-end} = \sqrt{\left\langle(\Vec{r}_N-\Vec{r}_1)^2\right\rangle}
\end{equation}

where $\langle....\rangle$ denotes an ensemble average.

Radius of gyration: One useful measure of chain size is the radius of gyration, denoted as $R_g$. 	For beads (of equal masses) connected by springs, the center of mass $\vec{r}_{cm}$ of the chain is defined as:
\begin{equation}\label{eq:18}
\vec{r}_{cm}=\frac{1}{N}\sum_{i=1}^{N}\vec{r}_i
\end{equation}
where $N$ is the number of beads and $\vec{r}_i$ is position vector of  
the $i^{th}$ bead. $R_g$ is defined as \cite{doi1988theory}
\begin{equation}\label{eq:19}
R_g = \sqrt{\frac{1}{N}\sum_{i=1}^{N} 
\left\langle\left| \vec{r}_i-\vec{r}_{cm}\right| ^2\right\rangle}
\end{equation}
 The $x$ component of $R_g$ is given as :
\begin{equation}\label{eq:20}
R_{g,x} = \sqrt{\frac{1}{N}\sum_{i=1}^{N} 
\left\langle\left({x}_i-{x}_{cm}\right)^2\right\rangle}
\end{equation}
where $x_i$ and  $x_{cm}$ denote the $x$-components of the position of the $i^{th}$ bead and the center of mass of the chain respectively. 
Similar expressions can be used for $y$ and $z$ components. In this study, $y$ is the flow direction. The $z$ and $x$ 
directions denote the shear-gradient and vorticity directions, 
respectively. Note, both $R_{end-to-end}$ and $R_{g}$ (and their $x$, $y$, $z$ components) can be selected as size measures. In this study, we have used the former for all cases, unless otherwise mentioned.

 Relaxation time: The autocorrelation function of end-to-end vector of the polymer chain provides an estimate of the relaxation time, as used in earlier studies \cite{kumar2023effectiveness}. The autocorrelation function of the end-to-end vector is defined as \cite{doi1988theory} 

\begin{equation}\label{eq:21}
C(t)=\langle\vec{R}(t)\cdot \vec{R}(0)\rangle
\end{equation}
where $\vec{R}=\vec{r}_N-\vec{r}_1$ denotes the end-to-end 
vector of the chain. The relaxation time is estimated by fitting the autocorrelation function to an exponential decay, as 
given by \cite{doi1988theory} :
\begin{equation}\label{eq:22}
\langle\vec{R}(t)\cdot\vec{R}(0)\rangle \cong \langle\vec{R}^2\rangle 
\exp\left(-\frac{t}{\tau}\right)
\end{equation}
where $\tau$ is the relaxation time, which is the characteristic time for the decay. Typically, the last $60-70\%$ is fitted to an exponential , as in previous studies \cite{jain2008effects}.    

The local dynamics within the chain is not completely described by the behavior of the end-to-end vector \cite{doi1988theory}. To obtain more of the local information, one other quantity is the autocorrelation of the bond vectors (used in some earlier studies, like \cite{saha2013explaining} and \cite{kumar2023effectiveness}). This is defined as \cite{peterson2001apparent}:

\begin{equation}\label{eq:22a}
\langle \vec{u}.\vec{u}\rangle =\frac{1}{N_s}\sum_{i=1,3,5...}^{N_s}exp(-t/\tau_i)
\end{equation}

where $\tau_i$ is the relaxation time of the $i^{th}$ mode and $N_s$ is the total number of springs (or \enquote{bonds}) in the chain. Here, the \enquote{mode} refers to the Rouse modes, as defined next. 

One effective measure of the local dynamics use the Normal modes, introduced by Rouse \cite{rouse1953theory} in his seminal work. These serve to decouple the equations of motion and provides the dynamics of a sub-segment of the chain.  The $i^{th}$ mode is defined as : 
\begin{equation}\label{eq:24}
\vec{q}_i(t) =\frac{1}{N}\sum_{n=1}^{N}\cos(\frac{in\pi}{N}) 
\vec{r}_n(t)
\end{equation}
where $\vec{q}_i(t)$ is the normal coordinate for the $i^{th}$ mode at 
any  time  $t$. The local dynamics is obtained by the autocorrelation functions of these normal modes.

Weissenberg number: This is the  product of the polymer chain relaxation time and the shear or extensional rate. Thus, $Wi=\dot{\epsilon}\tau$ for extensional flow and $Wi=\dot{\gamma}\tau$ for shear flow.

Strain: It is the dimensionless product of the shear or extensional rate and the simulation time. This is used instead of simulation time in all startup simulations, whenever needed.

\section{RESULTS AND DISCUSSION}
\subsection{DPD simulations in equilibrium}
Next, we assess the flexibility of DPD to access various regimes of polymer dynamics. Towards this, we will perform simulations at equilibrium (no flow). In all simulations, a polymer chain (with varying springs or Kuhn steps) is immersed in a solvent bath (sufficiently large). The typical default values of various parameters are taken as in the earlier study \cite{kumar2023effectiveness}. However, in this study, we perform simulations to assess the effect of the parameters, with an aim to understand the flexibility of DPD technique in modeling different regimes of polymer dynamics. The same will be investigated with a particular focus on polymer rheology (as studied in the latter half of this manuscript). Each parameter selected here is varied systematically across different chain lengths and its effects are studied on chain size and dynamics (including local dynamics through normal modes).  
\subsubsection{Effect of $K$}
First, we study whether variations in $K$ can change the dynamic (or size) regime of the polymer chain. In the earlier study \cite{kumar2023effectiveness}, the value of $K$ was varied in the range $10^3 - 10^4$, with an aim of keeping it sufficiently stiff (to mimic a Kuhn step) yet computationally feasible, since the time step size requirements rise steeply on increasing $K$. $K =5000$ was selected as the best compromise, which remains our default in this article. In a recent article using BD simulations of the bead-rod chain \cite{krishna2024analysis}, significant effects were observed while varying spring stiffness. Thus, to understand its implications completely in DPD, $K$ is systematically reduced from $5000$ to $25$ in this study, keeping the other parameters at their default values ($b =0.85$, $a_{ij} =25$  for all pairs). Fig. 1(a) shows the coil size variation with chain length, for different $K$. The inset shows the variation of the relaxation time, whereas Fig. 1(b) shows the scaling of the mode relaxation times for a chain of $60$ springs. Quite surprisingly, the chain size scaling exponent is $0.6$ for all values of $K$ (even the values overlap), suggesting negligible effects on the overall size. Note, this is when $K$ varies over two orders of magnitude. However, the bond length distribution with $K =25$ shows a large width, indicating that each spring for $K =25$ may not be behaving as a rigid rod (i.e. Kuhn step). This bond length distribution is shown in the inset of Fig. 1(b). Note how the distribution becomes noticeably wider for $K =25$.
\begin{figure} [h]
    
    \centering
    
    %\subfigure[]
    {
        \includegraphics[width= 3.7in,height=3.1in]{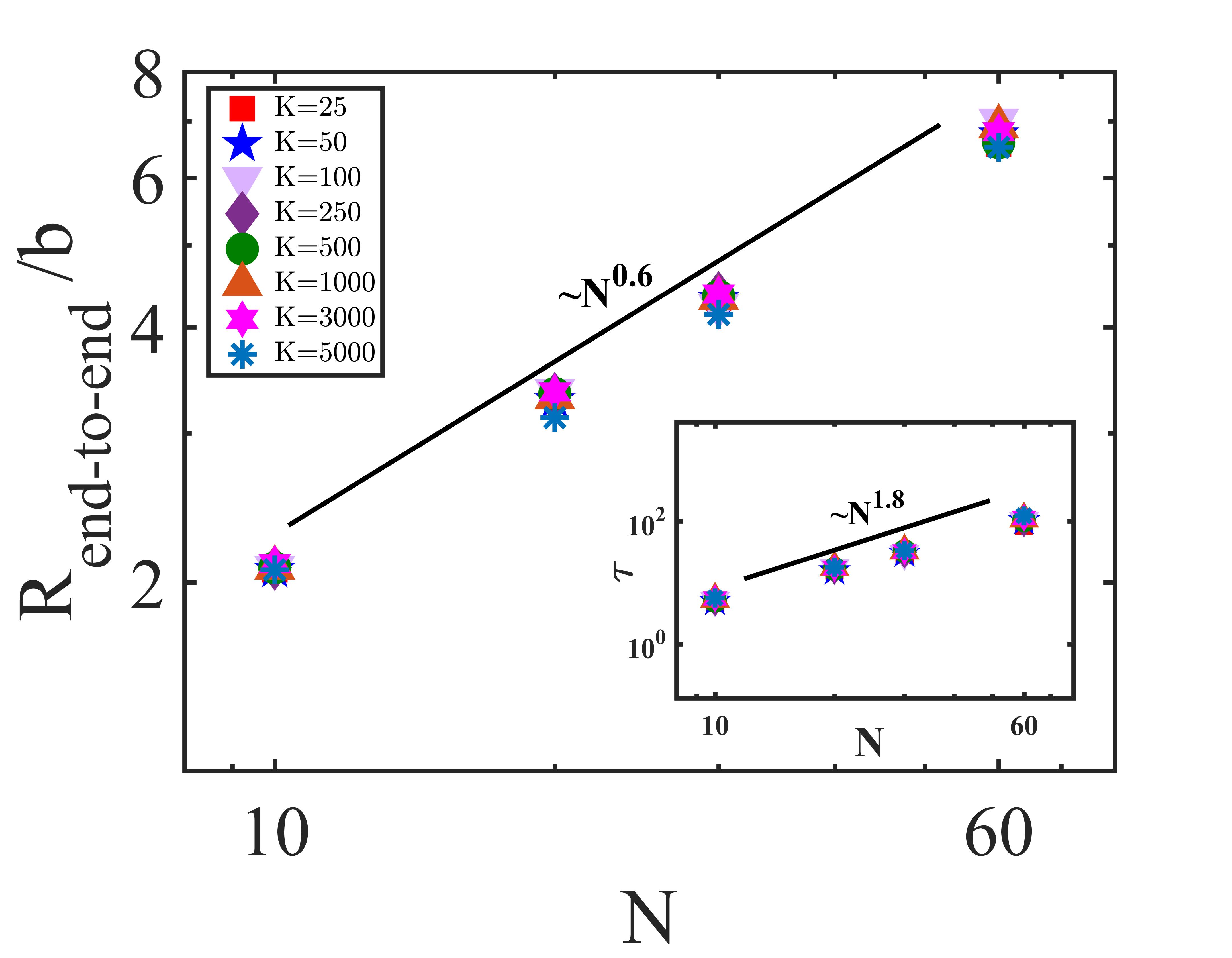}
        \put (-175,190){($a$)}
    }

   % \subfigure[]
    {
        \includegraphics[width= 3.7in,height=3.1in]{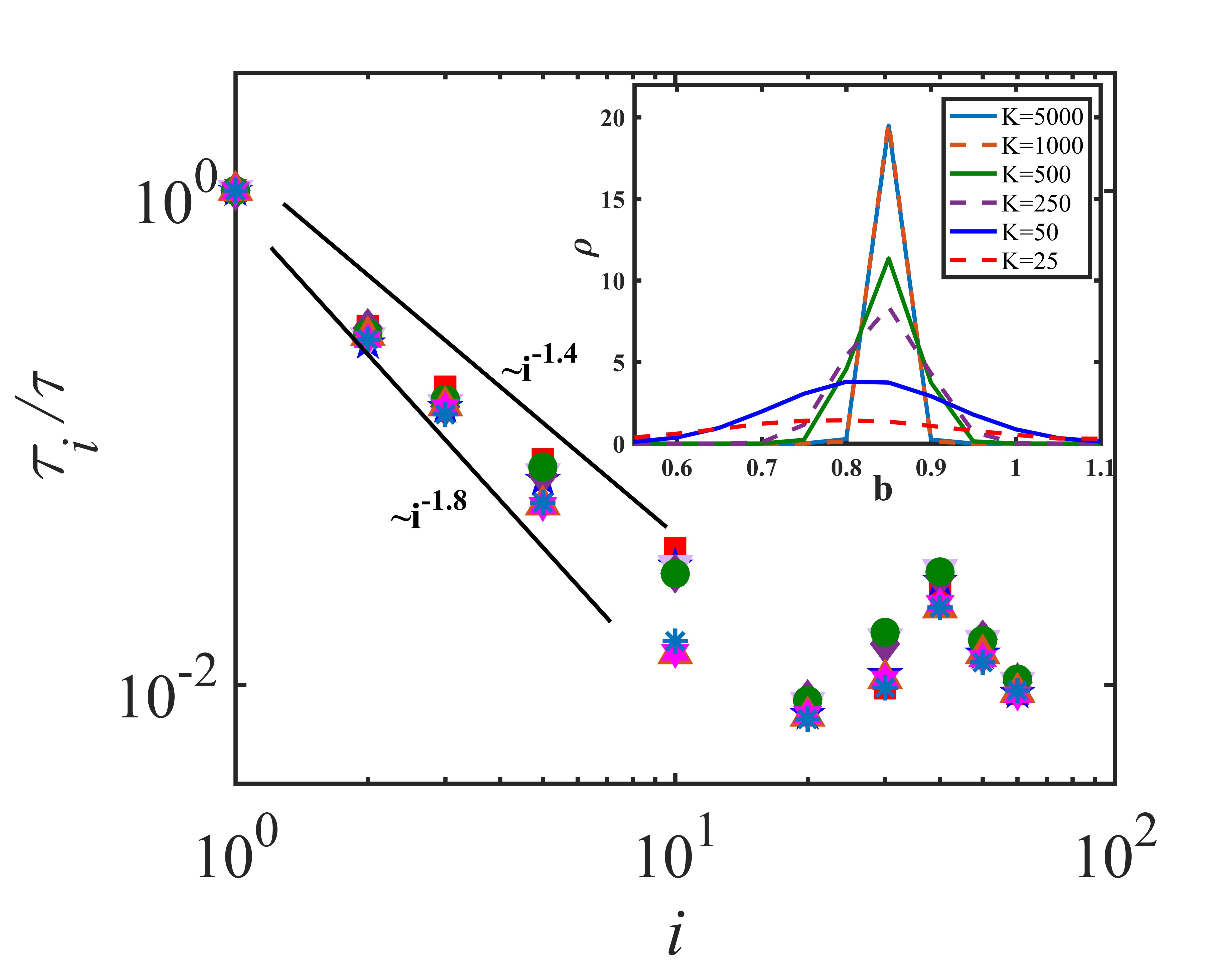}
         \put (-175,190){($b$)}
       % \label{}
    }

    \caption{{\footnotesize (a) Variation of $R_{end-to-end}$ and relaxation time $\tau$ (inset) with number of springs $N$, for different values of $K$. (b) Relaxation times $\tau_{i}$ of the $i^{th}$ mode versus $i$ for different values of $K$ for a chain of $60$ springs. The inset shows the bond length distributions of the chain for $K=5000$ (solid line) and $K=25$ (dashed line). The points represent the values from the DPD simulations, and the corresponding lines are the power-law fits of the results.}}
    \label{}
\end{figure}

% \begin{figure}  
% \centering
 %	\includegraphics[width=3.6in,height=3.3in]{allimages/1a.eps} 
% 	  \caption {Variation of $R_{end-to-end}$  and relaxation time $\tau$ with number of springs $N$, for different values of $K$. The points represent the values from the DPD simulations, and the corresponding lines are the power-law fits of the results.}

 %\end{figure}

The variation of the relaxation times (inset to Fig. 1(a)) also show negligible effects of $K$, in the range of $N$ studied here. Interestingly, some minor variations are observed at the largest chain length of $N=60$. Also, as discussed, the spring length distributions are not sharply peaked for lower values of $K$. Mostly thus, all practical (and consistent) $K$ values would yield similar values and similar scaling laws for the relaxation times. Thus, only the regime, where both HI and EV are dominant, can be accessible ($\tau \propto N^{1.8}$ and $R_{g} \propto N^{0.6}$) by varying $K$ within practical limits. The same conclusion is obtained from Fig. 1(b) showing the mode relaxation times. The relaxation times of various modes vary negligibly with varying $K$. The scaling regime suggests both HI and EV being dominant, with a saturation obtained for the higher modes (beyond $10^{th}$ mode). This has been observed and discussed in detail in the earlier study \cite{kumar2023effectiveness}, which is also consistent with experimental observations. Thus, since K variation does not enable access to other regimes, $K =5000$ would be suggested for all future similar DPD investigations, where a bead-rod chain is desired. Note, in the relaxation time spectrum a different scaling of about -1.4 is observed for the lower values of $K$. However, as discussed, the spring becomes softer at these $K$ values and should be avoided. 

% \begin{figure}  
 %\centering
 %	\includegraphics[width=3.6in,height=3.3in]{allimages/1b.eps} 
 %	  \caption {Relaxation times $\tau_{i}$ of the $i^{th}$ mode versus $i$ for different values of $K$ for chain of 60 springs. Bond length distribution with $K$, in inset. The points represent the values from the DPD simulations, and the corresponding lines are the power-law fits of the results.}
 %\end{figure}

\subsubsection{Effect of $a_{pp}$}

Next, we study the effects of $a_{pp}$, which controls the interaction between beads on the polymer chain. For all these studies, the other parameters are kept at their default values ($a_{ss} = a_{sp} = 25, K = 5000, b = 0.85$). As evident, the value of $a_{pp}$ will control the chain size, since it fixes the level of interaction within the polymer chain relative to other interactions (kept at $a_{ij} = 25$, which is typical for DPD simulations). However, the effects of $a_{pp}$ on the chain dynamics, especially local dynamics, has never been investigated, to the best of our knowledge. Here, we show the effects of $a_{pp}$ on chain size (Fig. 2(a)) and dynamics (Fig. 2(b)) (for local dynamics). The inset to Fig. 2(b) shows the variation of relaxation times with chain size. Our results suggest that $a_{pp}$ is an important parameter that allows us to access some regimes.

%\begin{figure}  
 %\centering
 %	\includegraphics[width=3.7in,height=3.4in]{allimages/2a.eps} 
 	  %\caption {Variation of $R_{end-to-end}$  with number of springs $N$, for different values of $a_{pp}$. The points represent the values from the DPD simulations, and the corresponding lines are the power-law fits of the results.}
 %\end{figure}

First, we examine the effect of $a_{pp}$ on the coil size (Fig. 2(a)). For computational limitations, we perform sufficiently long simulations on chains with at most 100 springs. One aspect from the results in Fig. 2(a) is the scaling laws for chain sizes in the range considered. Clearly, poor solvent behavior is obtained for low $a_{pp} < 20$. The exponent shows theta to good solvent behavior for $a_{pp} \geq 20$ (as indicated by the scaling laws shown by solid lines). The other aspect, not directly visible, are the values of the coil size for different $a_{pp}$ and $N$. Note that the spring equilibrium length (alternatively, length of a kuhn step for stiff springs) is $b=0.85$ in all cases. For this $b$, the coil size nearly matches that for an ideal random walk for $a_{pp}=55$ (shown in inset). $a_{pp}=25$ (default for DPD simulations) predicts slightly smaller coils, whereas $a_{pp}=100$ predicts swollen coils (relative to theta). Note, for this analysis, we used the larger chain sizes ($N \geq$ 30) and neglected the small chains, owing to the fact that EV effects may be negligible in smaller chains. Thus, from Fig. 2(a), $a_{pp}=55$ can be used to model theta coils in simulations. Swollen coils (good solvent conditions) may be obtained for larger values of $a_{pp}$.
\begin{figure} [h]
    
    \centering
    
    %\subfigure[]
    {
        \includegraphics[width= 3.7in,height=3.1in]{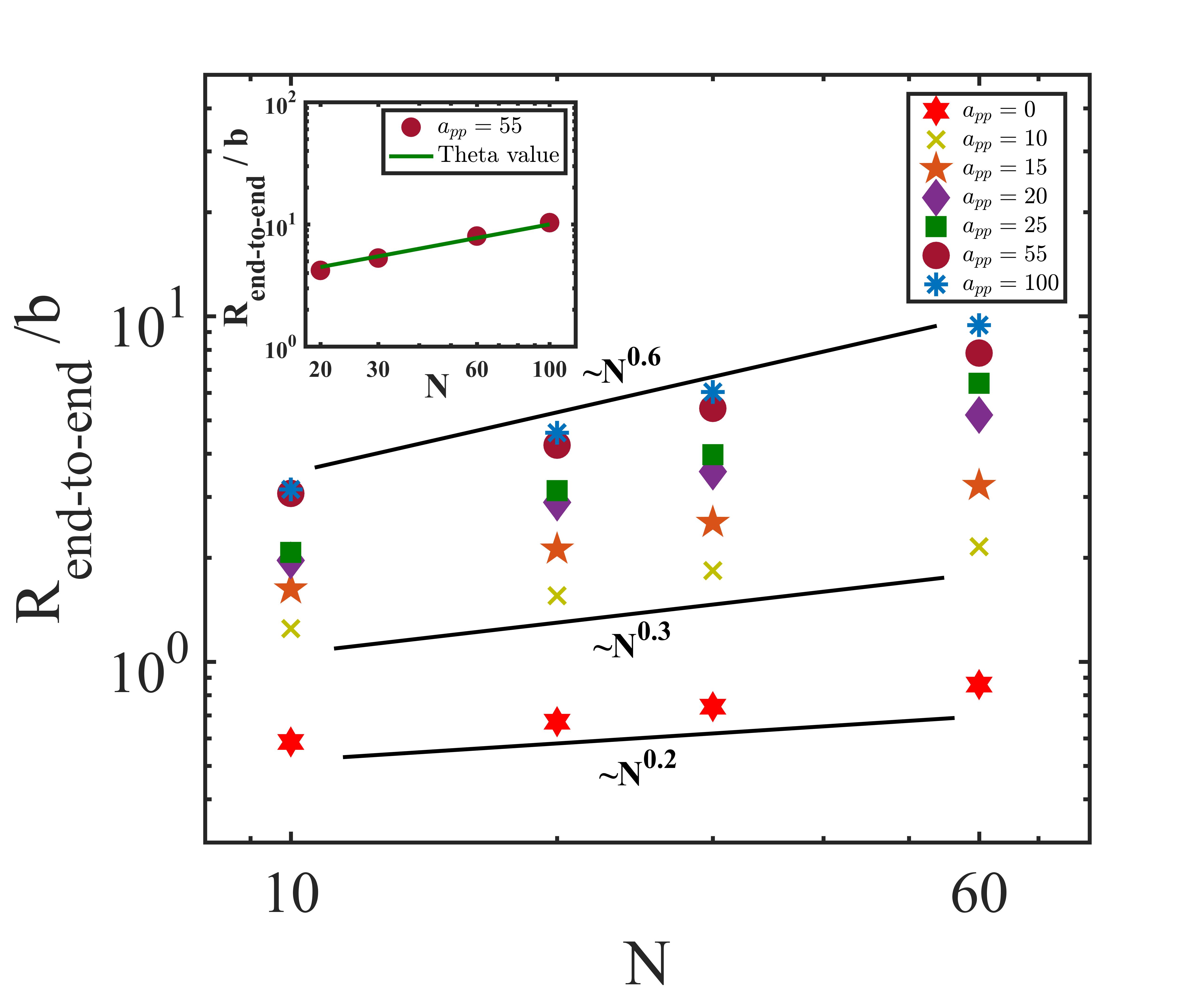}
        \put (-125,190){($a$)}
      %  \label{}
    }

   % \subfigure[]
    {
        \includegraphics[width= 3.7in,height=3.1in]{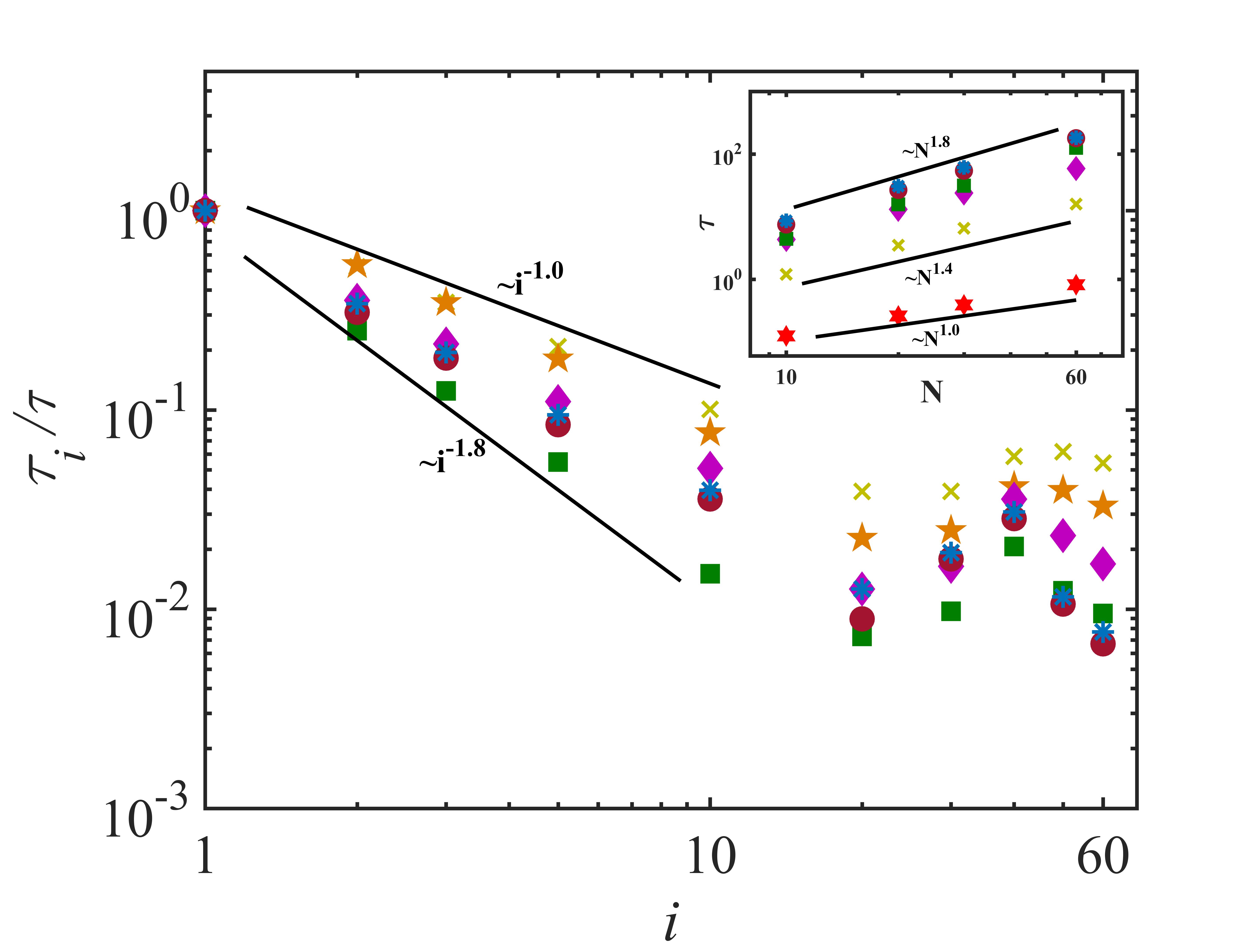}
        \put (-145,190){($b$)}
     %   \label{}
    }

    \caption{{\footnotesize (a) Variation of $R_{end-to-end}$ with number of springs $N$, for different values of $a_{pp}$. The inset shows the success of $a_{pp}=55$ in reproducing theta coils. (b) Relaxation times $\tau_{i}$ of the $i^{th}$ mode versus $i$ for a chain of $60$ springs. The inset shows the variation of the relaxation time $\tau$ with number of springs $N$, for different values of $a_{pp}$. The points represent the values from the DPD simulations, and the corresponding lines are the power-law fits of the results.}}
    \label{fig:E2E3}
\end{figure}
%\begin{figure}  
 %\centering
 %	\includegraphics[width=3.6in,height=3.3in]{allimages/2b.eps} 
 %	  \caption {Relaxation times $\tau_{i}$ of the $i^{th}$ mode  versus $i$ for chain of $60$ springs and relaxation time $\tau$ with number of springs $N$, for different values of $a_{pp}$ . The points represent the values from the DPD simulations, and the corresponding lines are the power-law fits of the results.}
 %\end{figure}

The chain dynamics for different $a_{pp}$ is shown in Fig. 2(b) (local dynamics through normal mode relaxation times of a chain of 60 springs) and its inset (relaxation times for different chain lengths). Note, the scaling law exponent is expected to vary between -1.5 (theta coil with HI) and -2 (no HI). When both EV and HI are dominant, an exponent of $-1.75$ to $-1.8$ is expected. The scaling law exponent is flatter than even -1.5 for $a_{pp} < 20$, which does not fall in any regime. For $a_{pp} \geq 20$, the exponents for the end-to-end relaxation times (inset of Fig. 2(b)) fall in the meaningful range of $-1.5$ to $-1.8$. However, note that the exponents are computed based on the longer chains. Similar exponents are observed for the mode relaxation times (Fig. 2(b)). Interestingly, a scaling exponent of about $-1.5$ is visible for the internal modes (other than the first mode), with a saturation reached beyond the $10^{th}$ mode, as observed with all DPD simulations, with $a_{pp} = 55$. For the classical value for DPD of 25 also used in the previous study, the local dynamics show an exponent of about $-1.8$ (both HI and EV dominant). Thus, $a_{pp} = 55$ provides the theta coil as well as predicts an exponent of $-1.5$ for the normal mode relaxation times, consistent with the Zimm predictions for a theta coil. Thus, clearly, the regime of a theta coil (no EV) and dominant HI ( mode relaxation time scale as $-1.5$) can be accessed by using $a_{pp} = 55$. This will be emphasized later as well when rheological simulations are performed. Good solvent conditions are expected for $a_{pp} > 55$ i.e. $100$ or above. For the default $a_{pp} = 25$, we end up in a regime where EV and HI effects are both present. 

\subsubsection{Effect of $b$}

In the previous section, we observed how the parameter $a_{pp}$ can be used to control the coil size. It was also evident that $a_{pp}$ has some effects on the chain dynamics. However, even though we can move from the poor solvent to theta and good solvent regimes using $a_{pp}$, it doesn't provide sufficient flexibility to traverse between dynamic regimes of dominant HI and negligible HI (Rouse-like scaling for relaxation times). For example, the user might need to access a regime of theta coil with negligible HI for some applications. Such regimes may be important for biological systems, like DNA chains that are known to exhibit negligible HI effects. Thus, to completely control the dynamic behavior of the chain, one other relevant parameter needs to be tuned. In this section, we show how the value of $b$ (spring length) allows us to access the Rouse-like and Zimm-like regimes.
%\begin{figure} [h]
    
 %   \centering
    
  %  \subfigure[]
   % {
    %    \includegraphics[width= 3.7in,height=3.1in]{allimages/3a.eps}
        %\label{}
    %}
   % \caption{{\footnotesize (a) Variation of $R_{end-to-end}$ with number of springs $N$, for different values of $b$.}}
   % \label{}
    %\subfigure[]
    %{
     %   \includegraphics[width= 3.7in,height=3.1in]{allimages/3b.eps}
      %  \label{}
    %}

    %\caption{{\footnotesize (b) Relaxation times $\tau_{i}$ of the $i^{th}$ mode versus $i$ for different values of $K$ for a chain of $60$ springs. The inset shows the bond length distributions of the chain for $K=5000$ (solid line) and $k=25$ (dashed line). The points represent the values from the DPD simulations, and the corresponding lines are the power-law fits of the results.}}
    %\label{}
%\end{figure} 
%\lipsum[1]
%\begin{figure}

    %\centering
    
   %\subfigure[] 
    
    %\includegraphics[width= 3.7in,height=3.1in]{allimages/3a.eps}
    %\put (-125,190){($a$)}
    
    %\caption{First subfigure.}
    %\label{}
  %\end{subfigure}
%\caption{\footnotesize (a) Variation of $R_{end-to-end}$ with number of springs $N$, for different values of $b$. The points represent the values from the DPD simulations, and the corresponding lines are the power-law fits of the results.}
%\label{fig:E5}
%\end{figure}
%%%%%%%% Continue figures %%%%%%%%
\begin{figure}[h]
%\ContinuedFloat
    
   %\subfigure[]
%    \centering
    
    \includegraphics[width= 7.0in,height=6.1in]{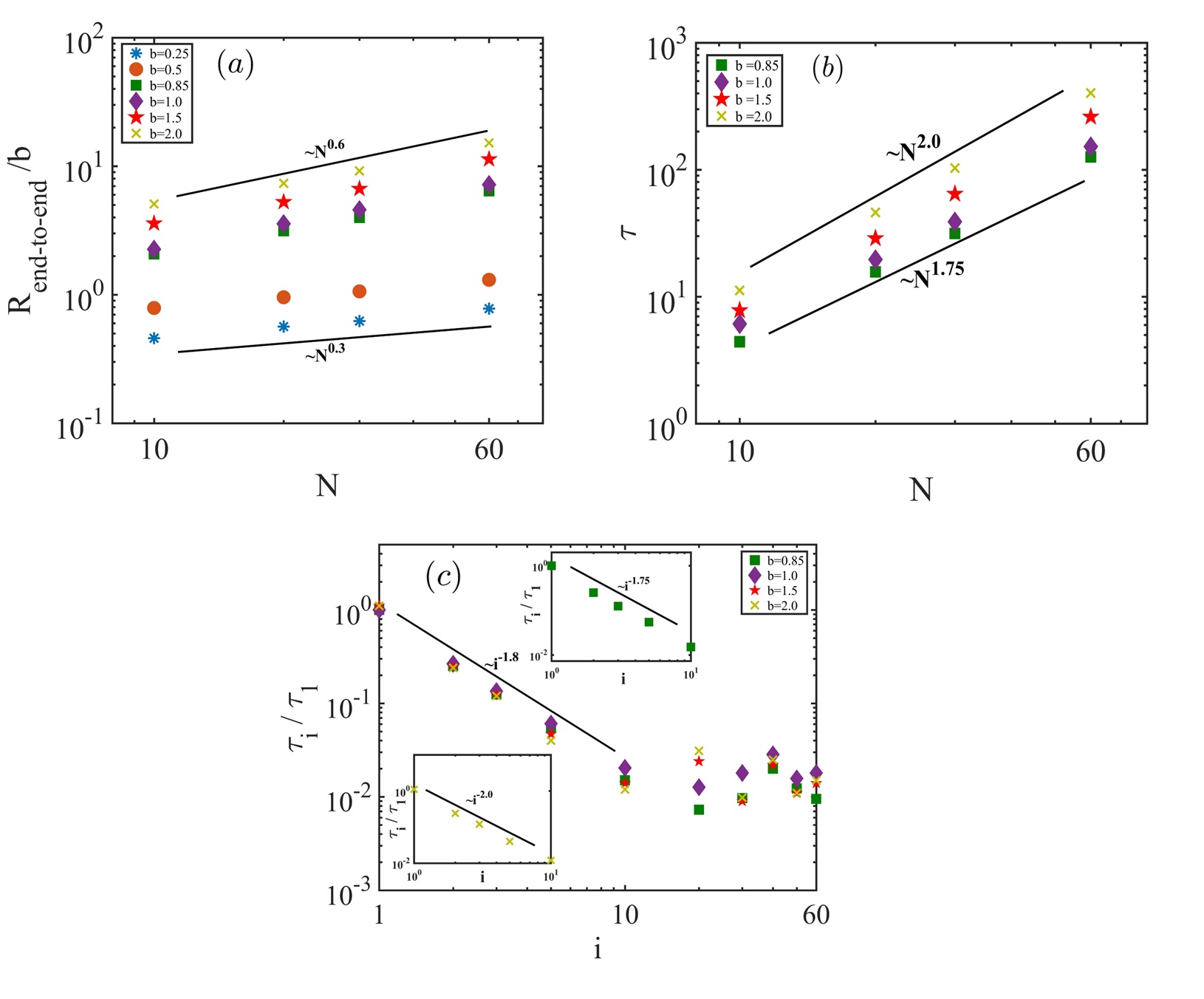}
     %\put (-135,190){($b$)}
    
    %\caption{Second subfigure.}
    %\label{}
% \end{subfigure}
    %\subfigure[]
%    \centering
    
    %\includegraphics[width= 3.7in,height=3.1in]{allimages/3c_final.eps}
    %\put (-190,190){($c$)}
    
    %\caption{Third subfigure.}
    %\label{}
  %\end{subfigure}

\caption{\footnotesize (a) Variation of $R_{end-to-end}$ with number of springs $N$, for different values of $b$. (b) Variation of relaxation time $\tau$ with number of springs $N$, for different values of $b$. (c) Relaxation times $\tau_{i}$ of the $i^{th}$ mode  versus $i$ for chain of $60$ springs for different values of $b$ . The points represent the values from the DPD simulations, and the corresponding lines are the power-law fits of the results.}
\label{}
\end{figure}
%\lipsum[2]
%\begin{figure} [h] 
 %\centering
 	%\includegraphics[width=3.6in,height=3.3in]{allimages/3a.eps} 
 	  %\caption {Variation of $R_{end-to-end}$  with number of springs $N$, for different values of $b$. The points represent the values from the DPD simulations, and the corresponding lines are the power-law fits of the results.}
 %   \label{fig:x cubed graph}
%\end{figure}
%\ref{fig:x cubed graph}
 Note, $b$ represents the Kuhn length for the polymer chain models considered in this study. The springs used are stiff, with the behavior matching with those used in the recent BD simulations for bead-rod models \cite{kumar2023effects}. It is well known that the effects of HI are negligible for polymer chains with larger Kuhn lengths (like DNA), whereas chains with relatively smaller Kuhn lengths (like polystyrene) exhibit dominant HI behavior. In our DPD setup, the Kuhn length can be adjusted through $b$, while keeping everything else constant. As we observe here, the value of $b$ does provide the necessary flexibility to access negligible HI and dominant HI regimes. This is in addition to $a_{pp}$, which allows us to traverse between theta and good solvent coil sizes.
%\begin{figure}  
 %\centering
 %	\includegraphics[width=3.6in,height=3.3in]{allimages/3b.eps} 
 %	  \caption {Variation of relaxation time $\tau$ with number of springs $N$, for different values of $b$. The points represent the values from the DPD simulations, and the corresponding lines are the power-law fits of the results.}
 %\end{figure}

Similar as the other factors, the value of $b$ is changed in DPD simulations to understand its effects on polymer size and dynamics. During these simulations, all other parameters were kept fixed at their default values ($a_{ss} = a_{pp} = a_{pp} = 25, K= 5000$). The scaling of chain size, end-to-end relaxation times, and normal mode relaxation times are shown in Figs. 3(a), (b), and (c), respectively, for $b$ values ranging from $0.25$ to $2.0$ (where the length scale is fixed by $r_{c}$, as discussed before). We observe a transition from the poor solvent regime at low $b$ ($0.25$ $-0.5$), to a good solvent at higher $b$ ($> 0.5$) (in Fig. 3(a)). Thus, for any value of $b$ $> 0.5$, for the default values of $a_{ij} = 25$, we obtain close to the good-solvent scaling exponent. Thus, for chain size aspects, any $b$ $> 0.5$ is preferable. Next, an analysis of the end-to-end relaxation time is presented in Fig. 3(b) for $0.85 \leq b \leq 2.0$ (omitting very low $b$ values where poor solvent regime is observed in Fig. 3(a)). Clearly, as anticipated in the discussion earlier, the Rouse-like regime of $\tau \propto N^{2}$ is observed for larger $b = 2$. A smooth transition is observed from an exponent of about $1.75$ (good solvent and HI) to $2$ (negligible HI). The same is also observed in the relaxation time of the normal modes (Fig. 3(c)), which further confirms the usefulness of b as a parameter to tune the level of HI. In the insets of Fig. 3(c), we show the scaling of the normal mode relaxation times for the two different regimes: $-1.75$ for HI (good solvent) and $-2.0$ for negligible HI. Thus, the factor $b$ provides us the flexibility to traverse across Rouse-like and Zimm-like regimes within the DPD framework (i.e. regimes of negligible HI and dominant HI).
%\begin{figure}  
 %\centering
 %	\includegraphics[width=3.6in,height=3.3in]{allimages/3c.eps} 
 %	  \caption {Relaxation times $\tau_{i}$ of the $i^{th}$ mode  versus $i$ for chain of $60$ springs for different values of $b$ . The points represent the values from the DPD simulations, and the corresponding lines are the power-law fits of the results.}
 %\end{figure}

\subsection{Summarizing DPD parameters for different rheologically important regimes}

Thus, we reveal two important parameters in this study that enable us to tune the coil size and level of HI. The first is $a_{pp}$, whose value can be tuned to obtain theta and swollen coils representing good solvent. Note, $a_{pp}$ affects the dynamics to some extent, enabling to move from scaling exponents of -1.75 (dominant HI + EV) to -1.5 (theta coil + dominant HI - Zimm theory). The spring length $b$ complements the effect of $a_{pp}$ by allowing the transition from negligible HI (Rouse scaling, -2·0) to dominant HI (-1.75, dominant HI + EV). Importantly, for any $b > 0.5$, the effect on coil size scaling law is negligible. Thus, we have nearly independent controls for the two important metrics for polymer dynamics simulations, -EV and HI. In Brownian Dynamics (BD) simulations, EV and HI can be set independently, thereby allowing us to suitably model the dynamics of any polymeric solution. In DPD, such independent control has never been revealed earlier, to the best of our knowledge. To summarize, we show how we can mimic a theta coil, with and without HI. To produce a thela coil, $a_{pp}=55$ is needed. If we use $b=2$, we obtain the case of negligible HI. For $b=1$, we arrive at dominant HI. This is indicated in Fig. 4, where the scaling of the normal mode relaxation times are shown for a chain of 100 springs. $a_{pp}=55$ is used throughout, $b=1$ and $b=2$ are used to show dominant HI and negligible HI, respectively. The scaling laws expected for these two regimes are obtained, as seen in Fig. 4. Note, it has already been shown that the theta size is obtained with $a_{pp}$.
\begin{figure}  
 \centering
 	\includegraphics[width=3.6in,height=3.3in]{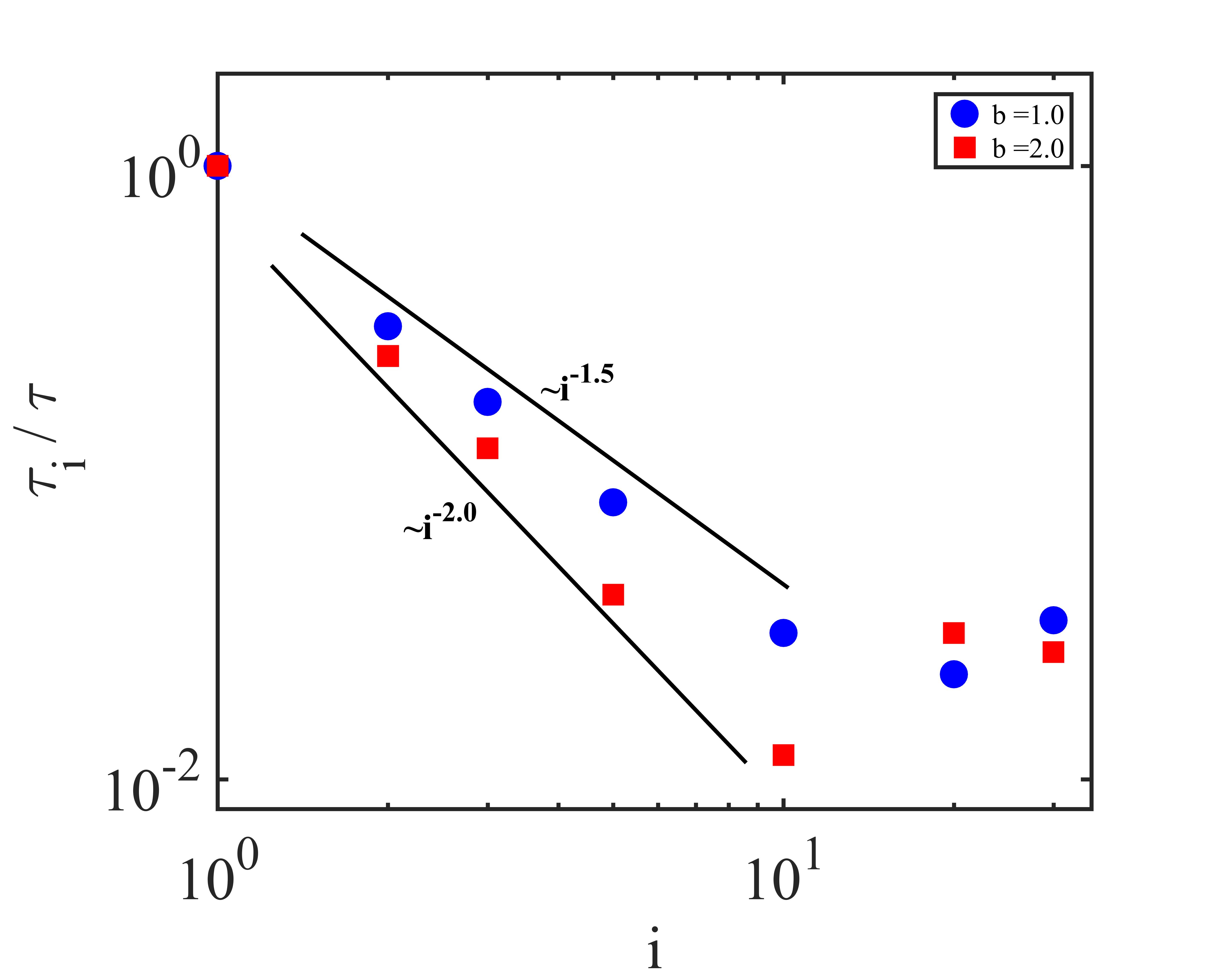} 
 	  \caption {Relaxation time spectrum for a theta coil ($a_{pp} = 55$) of a polymer chain of $100$ springs for two values of $b$, showing systems of dominant HI ($b = 1.0$) and negligible HI ($b=2.0$). The points represent the values from the DPD simulations, and the corresponding lines are the power-law fits of the results.}
 \end{figure}

\subsection{ Performance of DPD for rheological predictions}
Thus, we have established the parameter space and necessary values to access the different rheologically important regimes. To summarize, $a_{pp}$ produces the theta coil, where $b > 0.5$ enables us to traverse from regimes of dominant HI ($b = 0.85$ or $1$) to negligible HI ($b \geq 2.0$). Note, these values are to be adjusted in conjunction with others fixed at default levels ($a_{ss}=a_{sp}=25$, $K=5000$). This flexibility is extremely important and hitherto unknown. This enables DPD to be used for different polymeric solutions, ranging from DNA (weak HI) to polystyrene solutions (dominant HI). Note, in BD simulations, EV and HI effects can be independently switched on and off. The existence of a similar flexibility was previously not known in DPD, to the best of our knowledge.

The biggest advantage of DPD over BD is expected for flow situations in complex geometries. BD simulations always require diffusion tensors for HI. Such expressions are available for simple cases (like unbounded polymer solutions) but can't be formulated for complicated geometries, which occur in many applications. DPD does not require such specialized tensors and remains generally applicable for any situation. The only question that remained is of the flexibility in terms of adding/removing HI and EV, similar to BD. This aspect is clearly shown in this study, as summarized in the previous sections.

Even though our equilibrium results are quite indicative, further validations are required since these simulations are expected to be applied in flow situations. These validations
comprise of simulations in simpler, rheologically relevant flow fields. In this study, we use uniaxial extension and steady shear to further validate and bolster our results. These also serve to provide the foundation to investigations of applications, particularly rheological ones. Thus, here we probe if these DPD parameters are also able to provide good predictions for rheological behavior in various regimes, apart from equilibrium. Note, equilibrium behavior is the metric to separate out different dynamical regimes. To keep computational requirements practical, we perform DPD simulations for a chain of 100 springs (or  \enquote{rods}, since these are stiff springs) in uniaxial extension and shear, in only one dynamical regime. Here, we select the theta coil and dominant HI as the regime for flow simulations and compare with the results of BD simulations, which have been used for such unbounded systems for nearly all past investigations. As discussed, $a_{pp}=55$ for theta coil and $b=1$ for dominant HI in DPD. The other parameters remained at their default values. To compare with BD simulations of an equivalently resolved chain, we need to fix parameters for EV and HI (within BD) such that equilibrium behavior compares well with DPD. The details of BD simulations and governing equations are given in several recent articles \cite{krishna2024analysis}. For HI, the value of $h^*$(parameter for HI) needs to be fixed such that the normal mode spectrum matched with DPD. For EV, since a theta coil is desired, we have two choices. No EV interactions and EV interactions with  \enquote{just right} LJ parameters would both produce a theta coil. For a good comparison, we have used both approaches in BD. In this first approach, no interactions occur between beads and thus a theta coil is obtained. In the second, LJ parameters for bead-bead interactions is tuned such that the coil size is same as the theta coil.

After a few trial runs, the following values of BD parameters are obtained: $h^* = 0.1, \sigma = 0.5$ and $\epsilon =0.1$ for the system with EV that produced a theta coil. We refer to this as \enquote{theta-EV}. A chain without any EV also produced a theta coil (referred to as \enquote{no EV}) where $h^* = 0.1$ is used without any inter-bead interactions. The comparison of the relaxation spectrum is given in Fig. 5. All the results agree well, but the theta-EV spectrum is closer to that from the DPD simulations. Note, this is not unexpected, since the bead-bead interactions are also retained in the DPD simulations. The coil sizes also agree well, as noted here and are shown in the inset of Fig. 5. Next, all these systems are exposed to similar flow fields.

\begin{figure}  
 \centering
 	\includegraphics[width=3.6in,height=3.3in]{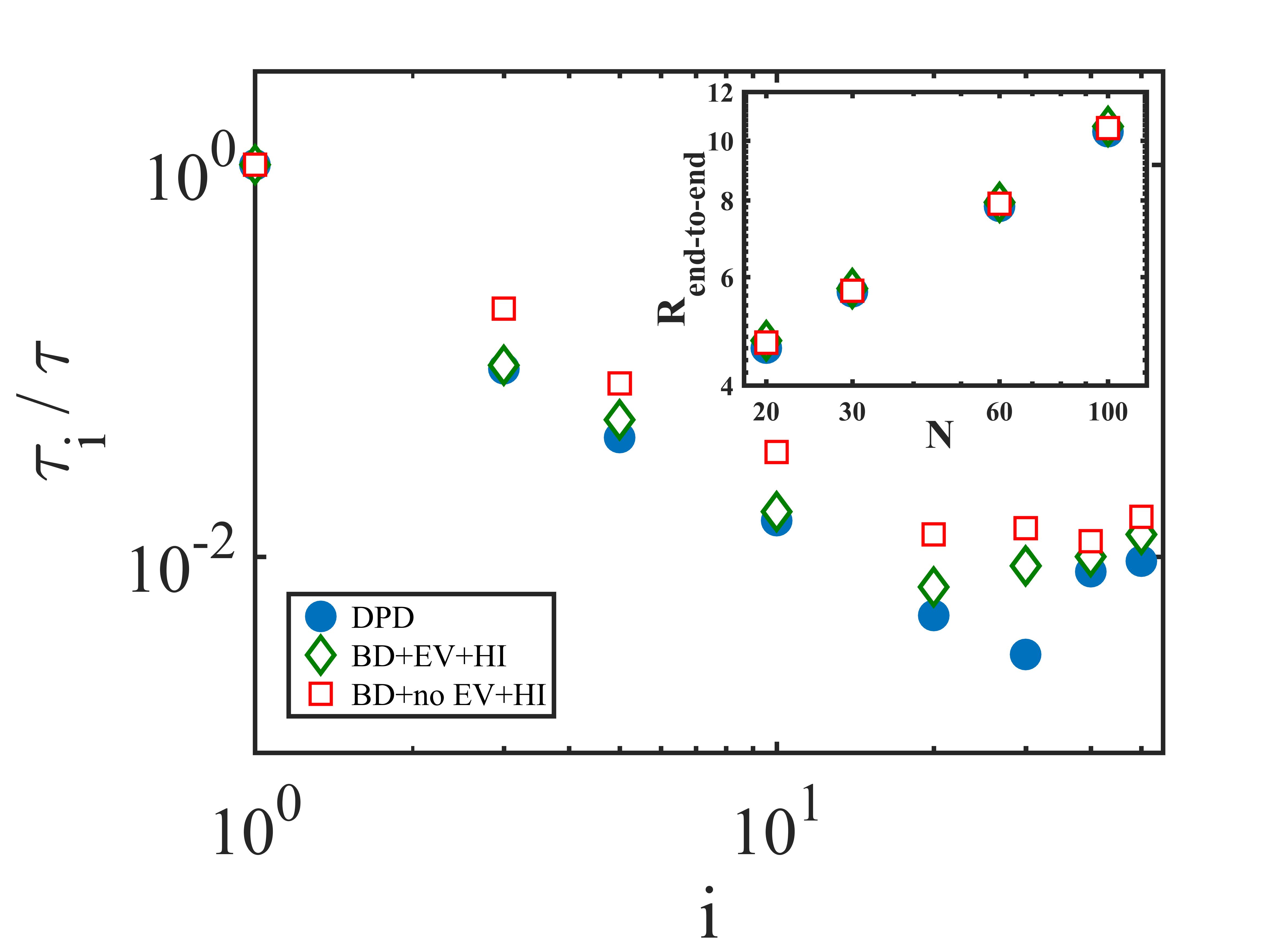} 
 	  \caption {Relaxation times $\tau_{i}$ of the $i^{th}$ mode  versus $i$ for chain of $100$ springs from DPD and BD simulations. The parameters for BD simulations use two different EV parameters to mimic a theta coil (marked as \enquote{no EV} and \enquote{EV}). For \enquote{EV}, the parameters are just right, such that a theta coil is reproduced. For details, see text.}
 \end{figure}
 \begin{figure}  
 \centering
 	\includegraphics[width=3.6in,height=3.3in]{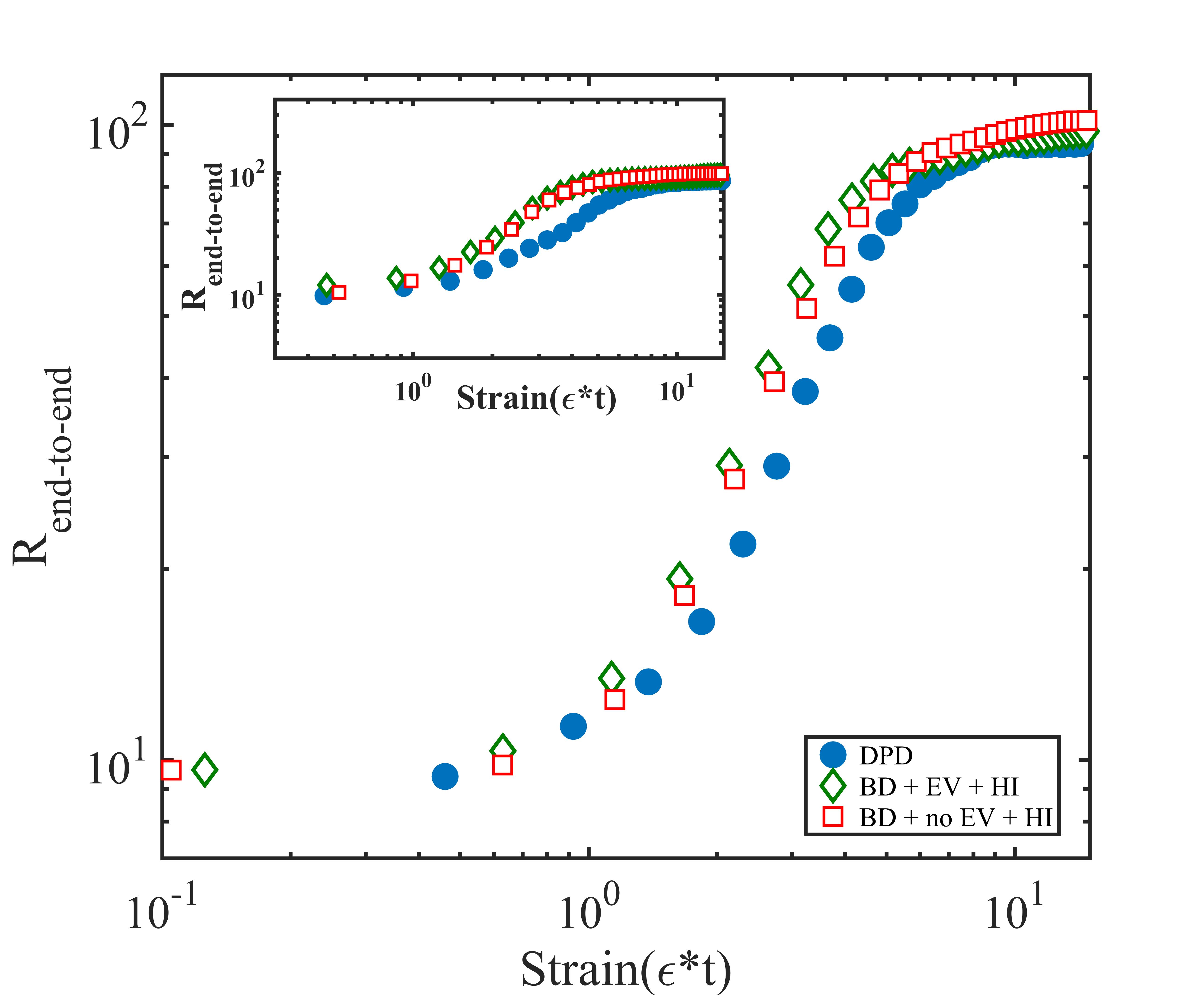} 
 	  \caption { Stretching dynamics of a polymer chain of $100$ springs in an uniaxial extensional flow of $Wi=20$ and $Wi=12.5$ (inset).
    Strain represents dimensionless time and $R_{end-to-end}$ represents the ensemble average end-to-end distance. Results are averaged over 100 and 75 cases for BD and DPD simulations, respectively.}
 \end{figure}
 \begin{figure}  
 \centering
 	\includegraphics[width=3.6in,height=3.3in]{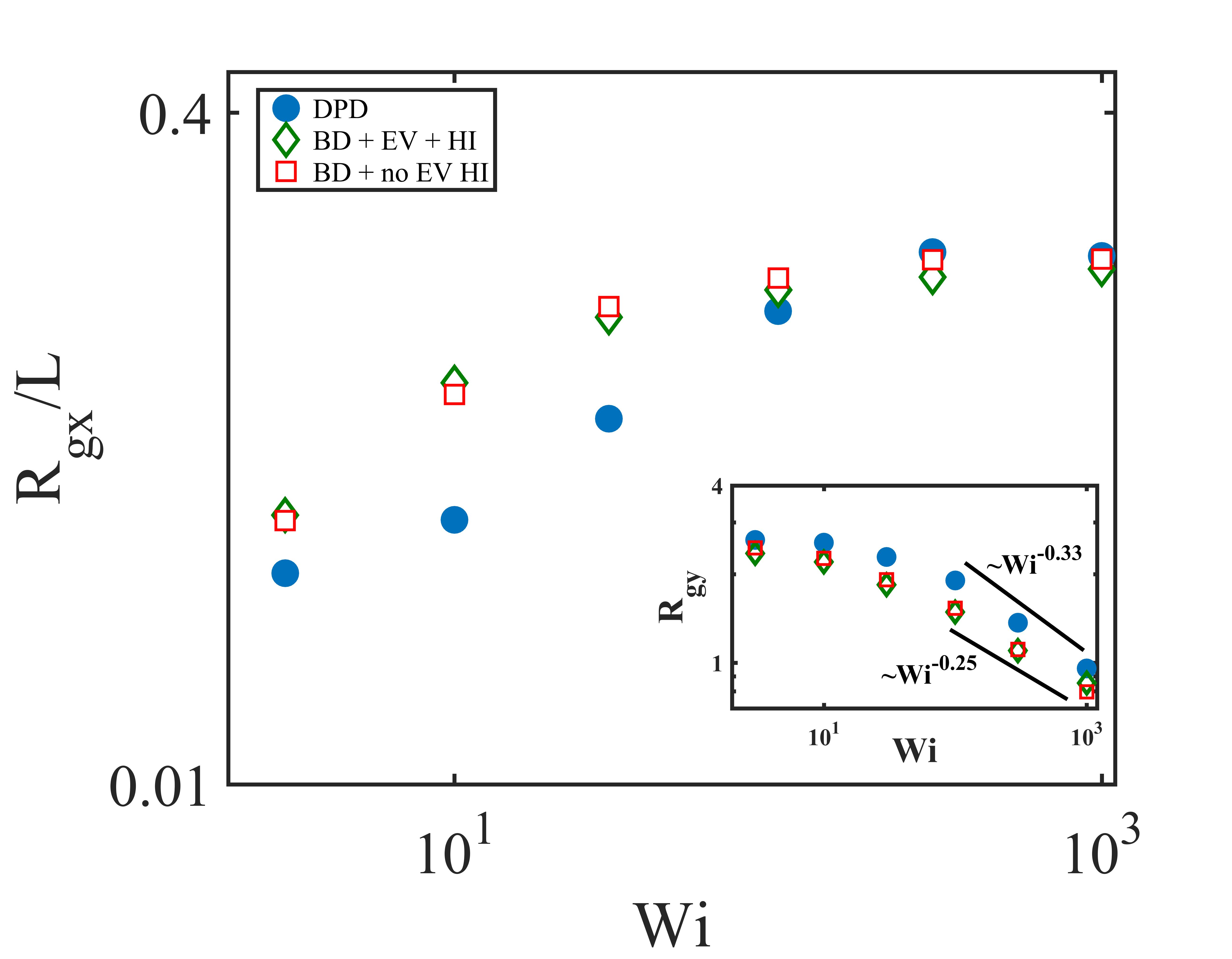} 
 	  \caption {Variation of the steady state radius of gyration along the flow direction ($R_{gx}$) normalized by contour length ($L$) with shear rate (denoted by $Wi$) for a polymer chain of $100$ springs in steady shear flow. The inset shows the variation of the radius of gyration along the gradient direction (coil thickness). The points are from DPD simulations and the lines indicate power-law fits of the same.}
 \end{figure}

The results for startup extensional flow are shown in Fig. 6. Note, we selected two different extensional rates as $Wi=12.5$ and $Wi=20$. Both are large enough to produce a final steady state that is nearly fully stretched. The dynamics of evolution of the end-to-end distance (averaged over 100 cases for BD simulations and 75 Cases for DPD simulations) is shown with strain ($\dot\epsilon t$, dimensionless time). For both cases, the DPD results show consistency with the BD predictions. For $Wi=12.5$  the DPD stretch predictions are slightly lower, but overall agrees well with BD. For the faster flow rate ($Wi=20$), the DPD results show excellent agreement with BD simulations. Note, these extensional flow calculations are performed in DPD in accordance with the method proposed earlier \cite{kumar2023effectiveness}. To the best of our knowledge, this is the first study using the method for simulating a polymer solution under extensional flow fields.

Lastly, we show comparisons of chain stretch in steady shear flow from BD and DPD simulations in Fig. 7. The shear flow is imposed in a similar manner as discussed in the earlier study with DPD in LAMMPS \cite{kumar2023effectiveness}. The steady state chain stretch (end-to-end distance) is obtained by imposing a shear flow on a chain and averaging over time. The trends are consistent with earlier observations for chains in shear \cite{dalal2012multiple, kumar2023effects}. The stretch increases till a saturation value is reached at strong shear rates. For all cases, the DPD results are in good agreement with BD simulations, especially for higher shear rates. However, the stretches are slightly lower for lower shear rates, but trends agree well throughout.
Interestingly, the scaling laws for the coil thickness shows some difference between BD and DPD. The BD simulations predict a scaling exponent of -1/4, whereas DPD simulations show -1/3. Note, this is a relatively shorter chain (100 Kuhn steps). Even with HI, for such chain lengths, the exponent of -1/4 is known from previous studies \cite{dalal2012multiple, kumar2023effects}. In another article \cite{saha2012tumbling}, it was theoretically argued that for longer
chains, in the presence of HI, the exponent should be close to -1/3. It is interesting to note that the DPD simulations show this exponent for a system where HI is dominant.

Thus, overall the DPD simulations agree well with the BD simulations for both extensional and shear flows considered here. The stretch agrees with BD predictions at high flow rates and the coil thickness shows the expected scaling laws in steady shear, for dominant HI. Hence, we have established the parameters that enable us to traverse between dynamical regimes, which is similar to BD in terms of flexibility. Also, predictions from our detailed DPD simulations in one regime (theta coil + dominant HI) agree well with those from BD simulations under similar conditions.

 %\begin{figure}  
 %\centering
 %	\includegraphics[width=3.6in,height=3.3in]{allimages/5b.eps} 
 %	  \caption {Variation of $R_{end-to-end}$  with strain for $w_{i}$ number $20$.}
 %\end{figure}

\medskip 
\section{Summary}

As discussed earlier, even though DPD is a mesoscale technique similar to BD simulations, the latter has been popular to study various aspects of polymer rheology even in recent times. This is due to the flexibility that BD offers in modelling various types of polymer solutions, because of the independent tuning of EV and HI. Being a general technique with beads exchanging momentum, DPD has some notable advantages, especially for flow complex geometries, where it may not be possible to formulate tensors for HI (for BD simulations). However, the absence of independent tunability of HI and EV makes it difficult to be used for any given polymer solution. Previously, this has not been shown to be possible within the framework of DPD. In this study, we highlight that this flexibility exists within DPD for polymer rheology. The spring length and the polymer bead interaction parameter provides two nearly independent tuning parameters for HI and EV, respectively. We show detailed
equilibrium polymer dynamics simulations to confirm this aspect, through coil size, relaxation times and normal mode spectrum. Next, once we establish the flexibility of DPD to model various dynamic regimes, the rheology of polymer solutions is compared with BD simulations. For this, the chain is subjected to extensional and steady shear flows. The regime of theta coil and dominant HI is selected for the rheological comparison with BD. Our results for chain stretch agree well with BD
simulations in both flow fields considered. Thus, it can be concluded that, with this aspect of tunability of DPD simulations and further validations with BD simulations for these standard flow fields, DPD can be used instead of BD for applications, especially those where BD may not be expected to perform well (or is difficult to formulate, like in complex geometries). 

\medskip
\begin{acknowledgments}
The authors express their gratitude for the kind assistance provided by
the generous support provided by the project sponsored by the Science $\&$
Engineering Research Board (SERB, Sanction No. CRG/2023/006625, Diary
No./Finance No. ANRF/F/722/2024-2025). We acknowledge the National Supercomputing Mission (NSM) for providing computing resources of 'PARAM Sanganak' at IIT Kanpur, which is implemented by C-DAC and supported by the Ministry of Electronics and Information Technology (MeitY) and Department of Science and Technology (DST), Government of India. Also, we would like to thank the computer center (www.iitk.ac.in/cc) at IIT Kanpur for providing the resources to carry out this work.
\end{acknowledgments}

%======================================	

\printendnotes
\newpage
%Submissions are not required to reflect the precise reference formatting of the journal (use of italics, bold, etc.); however, it is important that all key elements of each reference are included.
\bibliography{main}{}

\begin{thebibliography}{28}%
\makeatletter
\providecommand \@ifxundefined [1]{%
 \@ifx{#1\undefined}
}%
\providecommand \@ifnum [1]{%
 \ifnum #1\expandafter \@firstoftwo
 \else \expandafter \@secondoftwo
 \fi
}%
\providecommand \@ifx [1]{%
 \ifx #1\expandafter \@firstoftwo
 \else \expandafter \@secondoftwo
 \fi
}%
\providecommand \natexlab [1]{#1}%
\providecommand \enquote  [1]{``#1''}%
\providecommand \bibnamefont  [1]{#1}%
\providecommand \bibfnamefont [1]{#1}%
\providecommand \citenamefont [1]{#1}%
\providecommand \href@noop [0]{\@secondoftwo}%
\providecommand \href [0]{\begingroup \@sanitize@url \@href}%
\providecommand \@href[1]{\@@startlink{#1}\@@href}%
\providecommand \@@href[1]{\endgroup#1\@@endlink}%
\providecommand \@sanitize@url [0]{\catcode `\\12\catcode `\$12\catcode `\&12\catcode `\#12\catcode `\^12\catcode `\_12\catcode `\%12\relax}%
\providecommand \@@startlink[1]{}%
\providecommand \@@endlink[0]{}%
\providecommand \url  [0]{\begingroup\@sanitize@url \@url }%
\providecommand \@url [1]{\endgroup\@href {#1}{\urlprefix }}%
\providecommand \urlprefix  [0]{URL }%
\providecommand \Eprint [0]{\href }%
\providecommand \doibase [0]{http://dx.doi.org/}%
\providecommand \selectlanguage [0]{\@gobble}%
\providecommand \bibinfo  [0]{\@secondoftwo}%
\providecommand \bibfield  [0]{\@secondoftwo}%
\providecommand \translation [1]{[#1]}%
\providecommand \BibitemOpen [0]{}%
\providecommand \bibitemStop [0]{}%
\providecommand \bibitemNoStop [0]{.\EOS\space}%
\providecommand \EOS [0]{\spacefactor3000\relax}%
\providecommand \BibitemShut  [1]{\csname bibitem#1\endcsname}%
\let\auto@bib@innerbib\@empty
%</preamble>
\bibitem [{\citenamefont {Rouse~Jr}(1953)}]{rouse1953theory}%
  \BibitemOpen
  \bibfield  {author} {\bibinfo {author} {\bibfnamefont {P.~E.}\ \bibnamefont {Rouse~Jr}},\ }\bibfield  {title} {\enquote {\bibinfo {title} {A theory of the linear viscoelastic properties of dilute solutions of coiling polymers},}\ }\href@noop {} {\bibfield  {journal} {\bibinfo  {journal} {The Journal of Chemical Physics}\ }\textbf {\bibinfo {volume} {21}},\ \bibinfo {pages} {1272--1280} (\bibinfo {year} {1953})}\BibitemShut {NoStop}%
\bibitem [{\citenamefont {Zimm}(1956)}]{zimm1956dynamics}%
  \BibitemOpen
  \bibfield  {author} {\bibinfo {author} {\bibfnamefont {B.~H.}\ \bibnamefont {Zimm}},\ }\bibfield  {title} {\enquote {\bibinfo {title} {Dynamics of polymer molecules in dilute solution: viscoelasticity, flow birefringence and dielectric loss},}\ }\href@noop {} {\bibfield  {journal} {\bibinfo  {journal} {The journal of chemical physics}\ }\textbf {\bibinfo {volume} {24}},\ \bibinfo {pages} {269--278} (\bibinfo {year} {1956})}\BibitemShut {NoStop}%
\bibitem [{\citenamefont {Hsieh}, \citenamefont {Li},\ and\ \citenamefont {Larson}(2003)}]{hsieh2003modeling}%
  \BibitemOpen
  \bibfield  {author} {\bibinfo {author} {\bibfnamefont {C.-C.}\ \bibnamefont {Hsieh}}, \bibinfo {author} {\bibfnamefont {L.}~\bibnamefont {Li}}, \ and\ \bibinfo {author} {\bibfnamefont {R.~G.}\ \bibnamefont {Larson}},\ }\bibfield  {title} {\enquote {\bibinfo {title} {Modeling hydrodynamic interaction in brownian dynamics: simulations of extensional flows of dilute solutions of dna and polystyrene},}\ }\href@noop {} {\bibfield  {journal} {\bibinfo  {journal} {Journal of non-newtonian fluid mechanics}\ }\textbf {\bibinfo {volume} {113}},\ \bibinfo {pages} {147--191} (\bibinfo {year} {2003})}\BibitemShut {NoStop}%
\bibitem [{\citenamefont {Hsieh}\ and\ \citenamefont {Larson}(2004)}]{hsieh2004modeling}%
  \BibitemOpen
  \bibfield  {author} {\bibinfo {author} {\bibfnamefont {C.-C.}\ \bibnamefont {Hsieh}}\ and\ \bibinfo {author} {\bibfnamefont {R.~G.}\ \bibnamefont {Larson}},\ }\bibfield  {title} {\enquote {\bibinfo {title} {Modeling hydrodynamic interaction in brownian dynamics: Simulations of extensional and shear flows of dilute solutions of high molecular weight polystyrene},}\ }\href@noop {} {\bibfield  {journal} {\bibinfo  {journal} {Journal of Rheology}\ }\textbf {\bibinfo {volume} {48}},\ \bibinfo {pages} {995--1021} (\bibinfo {year} {2004})}\BibitemShut {NoStop}%
\bibitem [{\citenamefont {Petera}\ and\ \citenamefont {Muthukumar}(1999)}]{petera1999brownian}%
  \BibitemOpen
  \bibfield  {author} {\bibinfo {author} {\bibfnamefont {D.}~\bibnamefont {Petera}}\ and\ \bibinfo {author} {\bibfnamefont {M.}~\bibnamefont {Muthukumar}},\ }\bibfield  {title} {\enquote {\bibinfo {title} {Brownian dynamics simulation of bead--rod chains under shear with hydrodynamic interaction},}\ }\href@noop {} {\bibfield  {journal} {\bibinfo  {journal} {The Journal of chemical physics}\ }\textbf {\bibinfo {volume} {111}},\ \bibinfo {pages} {7614--7623} (\bibinfo {year} {1999})}\BibitemShut {NoStop}%
\bibitem [{\citenamefont {Hur}, \citenamefont {Shaqfeh},\ and\ \citenamefont {Larson}(2000)}]{hur2000brownian}%
  \BibitemOpen
  \bibfield  {author} {\bibinfo {author} {\bibfnamefont {J.~S.}\ \bibnamefont {Hur}}, \bibinfo {author} {\bibfnamefont {E.~S.}\ \bibnamefont {Shaqfeh}}, \ and\ \bibinfo {author} {\bibfnamefont {R.~G.}\ \bibnamefont {Larson}},\ }\bibfield  {title} {\enquote {\bibinfo {title} {Brownian dynamics simulations of single dna molecules in shear flow},}\ }\href@noop {} {\bibfield  {journal} {\bibinfo  {journal} {Journal of Rheology}\ }\textbf {\bibinfo {volume} {44}},\ \bibinfo {pages} {713--742} (\bibinfo {year} {2000})}\BibitemShut {NoStop}%
\bibitem [{\citenamefont {Larson}(2005)}]{larson2005rheology}%
  \BibitemOpen
  \bibfield  {author} {\bibinfo {author} {\bibfnamefont {R.~G.}\ \bibnamefont {Larson}},\ }\bibfield  {title} {\enquote {\bibinfo {title} {The rheology of dilute solutions of flexible polymers: Progress and problems},}\ }\href@noop {} {\bibfield  {journal} {\bibinfo  {journal} {Journal of Rheology}\ }\textbf {\bibinfo {volume} {49}},\ \bibinfo {pages} {1--70} (\bibinfo {year} {2005})}\BibitemShut {NoStop}%
\bibitem [{\citenamefont {Dalal}, \citenamefont {Hoda},\ and\ \citenamefont {Larson}(2012)}]{dalal2012multiple}%
  \BibitemOpen
  \bibfield  {author} {\bibinfo {author} {\bibfnamefont {I.~S.}\ \bibnamefont {Dalal}}, \bibinfo {author} {\bibfnamefont {N.}~\bibnamefont {Hoda}}, \ and\ \bibinfo {author} {\bibfnamefont {R.~G.}\ \bibnamefont {Larson}},\ }\bibfield  {title} {\enquote {\bibinfo {title} {Multiple regimes of deformation in shearing flow of isolated polymers},}\ }\href@noop {} {\bibfield  {journal} {\bibinfo  {journal} {Journal of Rheology}\ }\textbf {\bibinfo {volume} {56}},\ \bibinfo {pages} {305--332} (\bibinfo {year} {2012})}\BibitemShut {NoStop}%
\bibitem [{\citenamefont {Kumar}\ and\ \citenamefont {Saha~Dalal}(2022)}]{kumar2022fraenkel}%
  \BibitemOpen
  \bibfield  {author} {\bibinfo {author} {\bibfnamefont {P.}~\bibnamefont {Kumar}}\ and\ \bibinfo {author} {\bibfnamefont {I.}~\bibnamefont {Saha~Dalal}},\ }\bibfield  {title} {\enquote {\bibinfo {title} {Fraenkel springs as an efficient approximation to rods for brownian dynamics simulations and modeling of polymer chains},}\ }\href@noop {} {\bibfield  {journal} {\bibinfo  {journal} {Macromolecular Theory and Simulations}\ }\textbf {\bibinfo {volume} {31}},\ \bibinfo {pages} {2200008} (\bibinfo {year} {2022})}\BibitemShut {NoStop}%
\bibitem [{\citenamefont {Kumar}\ and\ \citenamefont {Dalal}(2023)}]{kumar2023effects}%
  \BibitemOpen
  \bibfield  {author} {\bibinfo {author} {\bibfnamefont {P.}~\bibnamefont {Kumar}}\ and\ \bibinfo {author} {\bibfnamefont {I.~S.}\ \bibnamefont {Dalal}},\ }\bibfield  {title} {\enquote {\bibinfo {title} {Effects of chain resolution on the configurational and rheological predictions from brownian dynamics simulations of an isolated polymer chain in flow},}\ }\href@noop {} {\bibfield  {journal} {\bibinfo  {journal} {Journal of Non-Newtonian Fluid Mechanics}\ }\textbf {\bibinfo {volume} {315}},\ \bibinfo {pages} {105017} (\bibinfo {year} {2023})}\BibitemShut {NoStop}%
\bibitem [{\citenamefont {Saha~Dalal}\ and\ \citenamefont {Larson}(2013)}]{saha2013explaining}%
  \BibitemOpen
  \bibfield  {author} {\bibinfo {author} {\bibfnamefont {I.}~\bibnamefont {Saha~Dalal}}\ and\ \bibinfo {author} {\bibfnamefont {R.~G.}\ \bibnamefont {Larson}},\ }\bibfield  {title} {\enquote {\bibinfo {title} {Explaining the absence of high-frequency viscoelastic relaxation modes of polymers in dilute solutions},}\ }\href@noop {} {\bibfield  {journal} {\bibinfo  {journal} {Macromolecules}\ }\textbf {\bibinfo {volume} {46}},\ \bibinfo {pages} {1981--1992} (\bibinfo {year} {2013})}\BibitemShut {NoStop}%
\bibitem [{\citenamefont {Krishna}, \citenamefont {Kumar},\ and\ \citenamefont {Saha~Dalal}(2024)}]{krishna2024analysis}%
  \BibitemOpen
  \bibfield  {author} {\bibinfo {author} {\bibfnamefont {S.}~\bibnamefont {Krishna}}, \bibinfo {author} {\bibfnamefont {P.}~\bibnamefont {Kumar}}, \ and\ \bibinfo {author} {\bibfnamefont {I.}~\bibnamefont {Saha~Dalal}},\ }\bibfield  {title} {\enquote {\bibinfo {title} {Analysis and estimation of time step sizes and stiffness parameters for efficient simulations of macromolecules with hydrodynamic interactions},}\ }\href@noop {} {\bibfield  {journal} {\bibinfo  {journal} {Physics of Fluids}\ }\textbf {\bibinfo {volume} {36}} (\bibinfo {year} {2024})}\BibitemShut {NoStop}%
\bibitem [{\citenamefont {Li}\ \emph {et~al.}(2013)\citenamefont {Li}, \citenamefont {Hu}, \citenamefont {Wang}, \citenamefont {Ma},\ and\ \citenamefont {Zhou}}]{li2013three}%
  \BibitemOpen
  \bibfield  {author} {\bibinfo {author} {\bibfnamefont {Z.}~\bibnamefont {Li}}, \bibinfo {author} {\bibfnamefont {G.-H.}\ \bibnamefont {Hu}}, \bibinfo {author} {\bibfnamefont {Z.-L.}\ \bibnamefont {Wang}}, \bibinfo {author} {\bibfnamefont {Y.-B.}\ \bibnamefont {Ma}}, \ and\ \bibinfo {author} {\bibfnamefont {Z.-W.}\ \bibnamefont {Zhou}},\ }\bibfield  {title} {\enquote {\bibinfo {title} {Three dimensional flow structures in a moving droplet on substrate: A dissipative particle dynamics study},}\ }\href@noop {} {\bibfield  {journal} {\bibinfo  {journal} {Physics of Fluids}\ }\textbf {\bibinfo {volume} {25}} (\bibinfo {year} {2013})}\BibitemShut {NoStop}%
\bibitem [{\citenamefont {Li}\ and\ \citenamefont {Drazer}(2008)}]{li2008hydrodynamic}%
  \BibitemOpen
  \bibfield  {author} {\bibinfo {author} {\bibfnamefont {Z.}~\bibnamefont {Li}}\ and\ \bibinfo {author} {\bibfnamefont {G.}~\bibnamefont {Drazer}},\ }\bibfield  {title} {\enquote {\bibinfo {title} {Hydrodynamic interactions in dissipative particle dynamics},}\ }\href@noop {} {\bibfield  {journal} {\bibinfo  {journal} {Physics of fluids}\ }\textbf {\bibinfo {volume} {20}} (\bibinfo {year} {2008})}\BibitemShut {NoStop}%
\bibitem [{\citenamefont {Liu}, \citenamefont {Meakin},\ and\ \citenamefont {Huang}(2007)}]{liu2007dissipative}%
  \BibitemOpen
  \bibfield  {author} {\bibinfo {author} {\bibfnamefont {M.}~\bibnamefont {Liu}}, \bibinfo {author} {\bibfnamefont {P.}~\bibnamefont {Meakin}}, \ and\ \bibinfo {author} {\bibfnamefont {H.}~\bibnamefont {Huang}},\ }\bibfield  {title} {\enquote {\bibinfo {title} {Dissipative particle dynamics simulation of multiphase fluid flow in microchannels and microchannel networks},}\ }\href@noop {} {\bibfield  {journal} {\bibinfo  {journal} {Physics of Fluids}\ }\textbf {\bibinfo {volume} {19}} (\bibinfo {year} {2007})}\BibitemShut {NoStop}%
\bibitem [{\citenamefont {Reddy}\ and\ \citenamefont {Abraham}(2009)}]{reddy2009dissipative}%
  \BibitemOpen
  \bibfield  {author} {\bibinfo {author} {\bibfnamefont {H.}~\bibnamefont {Reddy}}\ and\ \bibinfo {author} {\bibfnamefont {J.}~\bibnamefont {Abraham}},\ }\bibfield  {title} {\enquote {\bibinfo {title} {Dissipative-particle dynamics simulations of flow over a stationary sphere in compliant channels},}\ }\href@noop {} {\bibfield  {journal} {\bibinfo  {journal} {Physics of Fluids}\ }\textbf {\bibinfo {volume} {21}} (\bibinfo {year} {2009})}\BibitemShut {NoStop}%
\bibitem [{\citenamefont {Huang}, \citenamefont {Wang},\ and\ \citenamefont {Laradji}(2006)}]{huang2006flow}%
  \BibitemOpen
  \bibfield  {author} {\bibinfo {author} {\bibfnamefont {J.}~\bibnamefont {Huang}}, \bibinfo {author} {\bibfnamefont {Y.}~\bibnamefont {Wang}}, \ and\ \bibinfo {author} {\bibfnamefont {M.}~\bibnamefont {Laradji}},\ }\bibfield  {title} {\enquote {\bibinfo {title} {Flow control by smart nanofluidic channels: a dissipative particle dynamics simulation},}\ }\href@noop {} {\bibfield  {journal} {\bibinfo  {journal} {Macromolecules}\ }\textbf {\bibinfo {volume} {39}},\ \bibinfo {pages} {5546--5554} (\bibinfo {year} {2006})}\BibitemShut {NoStop}%
\bibitem [{\citenamefont {Wijmans}\ and\ \citenamefont {Smit}(2002)}]{wijmans2002simulating}%
  \BibitemOpen
  \bibfield  {author} {\bibinfo {author} {\bibfnamefont {C.}~\bibnamefont {Wijmans}}\ and\ \bibinfo {author} {\bibfnamefont {B.}~\bibnamefont {Smit}},\ }\bibfield  {title} {\enquote {\bibinfo {title} {Simulating tethered polymer layers in shear flow with the dissipative particle dynamics technique},}\ }\href@noop {} {\bibfield  {journal} {\bibinfo  {journal} {Macromolecules}\ }\textbf {\bibinfo {volume} {35}},\ \bibinfo {pages} {7138--7148} (\bibinfo {year} {2002})}\BibitemShut {NoStop}%
\bibitem [{\citenamefont {Jiang}\ \emph {et~al.}(2007)\citenamefont {Jiang}, \citenamefont {Huang}, \citenamefont {Wang},\ and\ \citenamefont {Laradji}}]{jiang2007hydrodynamic}%
  \BibitemOpen
  \bibfield  {author} {\bibinfo {author} {\bibfnamefont {W.}~\bibnamefont {Jiang}}, \bibinfo {author} {\bibfnamefont {J.}~\bibnamefont {Huang}}, \bibinfo {author} {\bibfnamefont {Y.}~\bibnamefont {Wang}}, \ and\ \bibinfo {author} {\bibfnamefont {M.}~\bibnamefont {Laradji}},\ }\bibfield  {title} {\enquote {\bibinfo {title} {Hydrodynamic interaction in polymer solutions simulated with dissipative particle dynamics},}\ }\href@noop {} {\bibfield  {journal} {\bibinfo  {journal} {The Journal of chemical physics}\ }\textbf {\bibinfo {volume} {126}} (\bibinfo {year} {2007})}\BibitemShut {NoStop}%
\bibitem [{\citenamefont {Kumar}\ \emph {et~al.}(2023)\citenamefont {Kumar}, \citenamefont {Jana}, \citenamefont {Shyam},\ and\ \citenamefont {Saha~Dalal}}]{kumar2023effectiveness}%
  \BibitemOpen
  \bibfield  {author} {\bibinfo {author} {\bibfnamefont {P.}~\bibnamefont {Kumar}}, \bibinfo {author} {\bibfnamefont {S.}~\bibnamefont {Jana}}, \bibinfo {author} {\bibfnamefont {H.}~\bibnamefont {Shyam}}, \ and\ \bibinfo {author} {\bibfnamefont {I.}~\bibnamefont {Saha~Dalal}},\ }\bibfield  {title} {\enquote {\bibinfo {title} {Effectiveness of dpd simulations to predict the dynamics of polymer chains in solutions at equilibrium and steady shear flows},}\ }\href@noop {} {\bibfield  {journal} {\bibinfo  {journal} {Macromolecular Theory and Simulations}\ }\textbf {\bibinfo {volume} {32}},\ \bibinfo {pages} {2300045} (\bibinfo {year} {2023})}\BibitemShut {NoStop}%
\bibitem [{\citenamefont {Kumar}\ \emph {et~al.}(2024)\citenamefont {Kumar}, \citenamefont {Krishna}, \citenamefont {Sharma},\ and\ \citenamefont {Saha~Dalal}}]{kumar2024effects}%
  \BibitemOpen
  \bibfield  {author} {\bibinfo {author} {\bibfnamefont {P.}~\bibnamefont {Kumar}}, \bibinfo {author} {\bibfnamefont {S.}~\bibnamefont {Krishna}}, \bibinfo {author} {\bibfnamefont {B.}~\bibnamefont {Sharma}}, \ and\ \bibinfo {author} {\bibfnamefont {I.}~\bibnamefont {Saha~Dalal}},\ }\bibfield  {title} {\enquote {\bibinfo {title} {Effects of chain resolution on the configurational and rheological predictions of dilute polymer solutions in flow fields with hydrodynamic interactions},}\ }\href@noop {} {\bibfield  {journal} {\bibinfo  {journal} {Physics of Fluids}\ }\textbf {\bibinfo {volume} {36}} (\bibinfo {year} {2024})}\BibitemShut {NoStop}%
\bibitem [{\citenamefont {Groot}\ and\ \citenamefont {Warren}(1997)}]{groot1997dissipative}%
  \BibitemOpen
  \bibfield  {author} {\bibinfo {author} {\bibfnamefont {R.~D.}\ \bibnamefont {Groot}}\ and\ \bibinfo {author} {\bibfnamefont {P.~B.}\ \bibnamefont {Warren}},\ }\bibfield  {title} {\enquote {\bibinfo {title} {Dissipative particle dynamics: Bridging the gap between atomistic and mesoscopic simulation},}\ }\href@noop {} {\bibfield  {journal} {\bibinfo  {journal} {The Journal of chemical physics}\ }\textbf {\bibinfo {volume} {107}},\ \bibinfo {pages} {4423--4435} (\bibinfo {year} {1997})}\BibitemShut {NoStop}%
\bibitem [{\citenamefont {Espanol}\ and\ \citenamefont {Warren}(1995)}]{espanol1995statistical}%
  \BibitemOpen
  \bibfield  {author} {\bibinfo {author} {\bibfnamefont {P.}~\bibnamefont {Espanol}}\ and\ \bibinfo {author} {\bibfnamefont {P.}~\bibnamefont {Warren}},\ }\bibfield  {title} {\enquote {\bibinfo {title} {Statistical mechanics of dissipative particle dynamics},}\ }\href@noop {} {\bibfield  {journal} {\bibinfo  {journal} {Europhysics letters}\ }\textbf {\bibinfo {volume} {30}},\ \bibinfo {pages} {191} (\bibinfo {year} {1995})}\BibitemShut {NoStop}%
\bibitem [{\citenamefont {Plimpton}(1995)}]{plimpton1995fast}%
  \BibitemOpen
  \bibfield  {author} {\bibinfo {author} {\bibfnamefont {S.}~\bibnamefont {Plimpton}},\ }\bibfield  {title} {\enquote {\bibinfo {title} {Fast parallel algorithms for short-range molecular dynamics},}\ }\href@noop {} {\bibfield  {journal} {\bibinfo  {journal} {Journal of computational physics}\ }\textbf {\bibinfo {volume} {117}},\ \bibinfo {pages} {1--19} (\bibinfo {year} {1995})}\BibitemShut {NoStop}%
\bibitem [{\citenamefont {Doi}\ and\ \citenamefont {Edwards}(1988)}]{doi1988theory}%
  \BibitemOpen
  \bibfield  {author} {\bibinfo {author} {\bibfnamefont {M.}~\bibnamefont {Doi}}\ and\ \bibinfo {author} {\bibfnamefont {S.~F.}\ \bibnamefont {Edwards}},\ }\href@noop {} {\emph {\bibinfo {title} {The theory of polymer dynamics}}},\ Vol.~\bibinfo {volume} {73}\ (\bibinfo  {publisher} {oxford university press},\ \bibinfo {year} {1988})\BibitemShut {NoStop}%
\bibitem [{\citenamefont {Jain}\ and\ \citenamefont {Larson}(2008)}]{jain2008effects}%
  \BibitemOpen
  \bibfield  {author} {\bibinfo {author} {\bibfnamefont {S.}~\bibnamefont {Jain}}\ and\ \bibinfo {author} {\bibfnamefont {R.~G.}\ \bibnamefont {Larson}},\ }\bibfield  {title} {\enquote {\bibinfo {title} {Effects of bending and torsional potentials on high-frequency viscoelasticity of dilute polymer solutions},}\ }\href@noop {} {\bibfield  {journal} {\bibinfo  {journal} {Macromolecules}\ }\textbf {\bibinfo {volume} {41}},\ \bibinfo {pages} {3692--3700} (\bibinfo {year} {2008})}\BibitemShut {NoStop}%
\bibitem [{\citenamefont {Peterson}\ \emph {et~al.}(2001)\citenamefont {Peterson}, \citenamefont {Echeverr{\'\i}a}, \citenamefont {Hahn}, \citenamefont {Strand},\ and\ \citenamefont {Schrag}}]{peterson2001apparent}%
  \BibitemOpen
  \bibfield  {author} {\bibinfo {author} {\bibfnamefont {S.~C.}\ \bibnamefont {Peterson}}, \bibinfo {author} {\bibfnamefont {I.}~\bibnamefont {Echeverr{\'\i}a}}, \bibinfo {author} {\bibfnamefont {S.~F.}\ \bibnamefont {Hahn}}, \bibinfo {author} {\bibfnamefont {D.~A.}\ \bibnamefont {Strand}}, \ and\ \bibinfo {author} {\bibfnamefont {J.~L.}\ \bibnamefont {Schrag}},\ }\bibfield  {title} {\enquote {\bibinfo {title} {Apparent relaxation-time spectrum cutoff in dilute polymer solutions: An effect of solvent dynamics},}\ }\href@noop {} {\bibfield  {journal} {\bibinfo  {journal} {Journal of Polymer Science Part B: Polymer Physics}\ }\textbf {\bibinfo {volume} {39}},\ \bibinfo {pages} {2860--2873} (\bibinfo {year} {2001})}\BibitemShut {NoStop}%
\bibitem [{\citenamefont {Saha~Dalal}\ \emph {et~al.}(2012)\citenamefont {Saha~Dalal}, \citenamefont {Albaugh}, \citenamefont {Hoda},\ and\ \citenamefont {Larson}}]{saha2012tumbling}%
  \BibitemOpen
  \bibfield  {author} {\bibinfo {author} {\bibfnamefont {I.}~\bibnamefont {Saha~Dalal}}, \bibinfo {author} {\bibfnamefont {A.}~\bibnamefont {Albaugh}}, \bibinfo {author} {\bibfnamefont {N.}~\bibnamefont {Hoda}}, \ and\ \bibinfo {author} {\bibfnamefont {R.~G.}\ \bibnamefont {Larson}},\ }\bibfield  {title} {\enquote {\bibinfo {title} {Tumbling and deformation of isolated polymer chains in shearing flow},}\ }\href@noop {} {\bibfield  {journal} {\bibinfo  {journal} {Macromolecules}\ }\textbf {\bibinfo {volume} {45}},\ \bibinfo {pages} {9493--9499} (\bibinfo {year} {2012})}\BibitemShut {NoStop}%
\end{thebibliography}%

\end{document}